\documentclass[preprint,authoryear,3p,twocolumn]{elsarticle}

\usepackage{amssymb}
\usepackage{amsmath}
\usepackage{natbib}
\usepackage{xcolor}
\usepackage{subfigure}
\usepackage{ctable}
\usepackage{makecell}
\usepackage{caption} 
\usepackage[shortlabels]{enumitem}
\usepackage{hyperref}
\usepackage{xspace}

\usepackage{xcolor}
\usepackage{aas_macros}

\newcommand{\NEW}[1]{{#1}}

\journal{Astronomy and Computing}

\begin{document}

\begin{frontmatter}



\title{DeepGhostBusters: Using Mask R-CNN to Detect and Mask Ghosting and Scattered-Light Artifacts from Optical Survey Images}

\author[label1,label2]{Dimitrios Tanoglidis\corref{label5}}
\cortext[label5]{Corresponding author; FERMILAB-PUB-21-374-AE}
\ead{dtanoglidis@uchicago.edu}
\author[label3]{Aleksandra \'Ciprijanovi\'c}
\author[label3,label2,label1]{Alex Drlica-Wagner}
\author[label3,label2,label1]{Brian Nord}
\author[label3]{Michael H.~L.~S. Wang}
\author[label2]{Ariel Jacob Amsellem}
\author[label1]{Kathryn Downey}
\author[label1,label5]{Sydney Jenkins}
\author[label3]{Diana Kafkes}
\author[label4]{Zhuoqi Zhang}

\address[label1]{Department of Astronomy and Astrophysics, University of Chicago, Chicago, IL 60637, USA}
\address[label2]{Kavli Institute for Cosmological Physics, University of Chicago, Chicago, IL 60637, USA}
\address[label3]{Fermi National Accelerator Laboratory, P. O. Box 500, Batavia, IL 60510, USA}
\address[label4]{University of Chicago, Chicago, IL 60637, USA}
\address[label6]{Department of Physics, Massachusetts Institute of Technology, 77 Massachusetts Ave., Cambridge, MA 02139, USA}

\begin{abstract}
Wide-field astronomical surveys are often affected by the presence of undesirable reflections (often known as ``ghosting artifacts'' or ``ghosts") and scattered-light artifacts. 
The identification and mitigation of these artifacts is important for rigorous astronomical analyses of faint and  low-surface-brightness systems. 
However, the identification of ghosts and scattered-light artifacts is challenging due to a) the complex morphology of these features and b) the large data volume of current and near-future surveys. 
In this work, we use images from the Dark Energy Survey (DES) to train, validate, and test a deep neural network (Mask R-CNN) to detect and localize ghosts and scattered-light artifacts. 
We find that the ability of the Mask R-CNN model to identify affected regions is superior to that of conventional algorithms and traditional convolutional neural networks methods. 
We propose that a multi-step pipeline combining Mask R-CNN segmentation with a classical CNN classifier provides a powerful technique for the automated detection of ghosting and scattered-light artifacts in current and near-future surveys.
\end{abstract}

\end{frontmatter}


\section{Introduction}
\label{sec: inroduction}

Wide-field photometric surveys at optical and near-infrared wavelengths have provided a wealth of astronomical information that has enabled a better understanding of the processes that govern the growth and evolution of the Universe and its contents. 
Near-future surveys, such as the Vera C.\ Rubin Observatory's Legacy Survey of Space and Time \citep[LSST;][]{Ivezic:2019}\footnote{\url{https://www.lsst.org/}}, will further expand our knowledge of the Universe by extending measurements to unprecedentedly faint astronomical systems.
Such surveys will produce terabytes of data each night and measure tens of billions of stars and galaxies.

Images collected by optical/near-infrared surveys often contain imaging artifacts caused by scattered and reflected light (commonly known as ``ghosting artifacts'' or ``ghosts") from bright astronomical sources. 
These image artifacts are an unavoidable feature of many optical systems.
The effective mitigation of ghosts and scattered-light artifacts, and the spurious brightness variations they introduce, is important for the detection and precise measurement of faint astronomical systems. 
In particular, since many ghosts cover a large image area with relatively low surface brightness, they constitute a significant source of contamination in studies of the low-surface-brightness Universe, a major goal of current and upcoming surveys \citep[e.g.,][]{Greco_2018,brough2020vera,sugata:2020,Tanoglidis_2021a}.

\NEW{Modern wide-field telescopes and instruments} greatly reduce the occurrence and intensity of ghosts and scattered-light artifacts by introducing light baffles, and high efficiency anti-reflective coatings on key optical surfaces. 
\NEW{Strict requirements on the number and intensity of ghosts and scattered-light artifacts were achieved during the construction of the Dark Energy Camera \citep[DECam;][]{DECam:2009,Flaugher_2015}, which has enabled state-of-the-art cosmological analyses with the Dark Energy Survey \citep[DES;][]{DES:2005,DES:2016,DES_Y1KP,DES_Y3KP}.}\footnote{\url{https://www.darkenergysurvey.org/}},
Other smaller surveys have implemented novel optical designs to mitigate the presence of ghosts and scattered-light artifacts \citep{Abraham:2014}. 

Despite these successful efforts, it is often impossible to completely remove ghosts and scattered-light artifacts. 
For example, the DES 3-year cosmology analyses masked ${\sim}3\%$ of the survey area around the brightest stars, and ${\sim}10\%$ of the survey area around fainter stars \citep{Sevilla-Noarbe:2021}. 
Additional mitigation steps that go beyond the original survey design requirements are particularly important for studies of low-surface-brightness systems.

The large datasets produced by surveys like DES make the rejection of these residual artifacts by visual inspection infeasible. The situation will become even more intractable in upcoming surveys, like LSST, which will collect $\sim 20$TB/night and $\sim 15$PB of data over its nominal 10-year survey.\footnote{\url{https://www.lsst.org/scientists/keynumbers}}
Furthermore, the deeper imaging of LSST will place even tighter requirements on low-surface-brightness artifacts \citep{LSST_book,brough2020vera}.

To mitigate residual ghosts and scattered-light artifacts, DES uses a predictive Ray-Tracing algorithm as the core of its detection process. This algorithm forward models the physical processes that lead to ghosting/scattered-light events \citep{Kent:2013}, such as the configuration of the telescope and camera optics, and the positions and brightnesses of known stars obtained from catalogs external to the survey (for a more detailed description of the Ray-Tracing algorithm, see \citealt{Kent:2013} and Sec.~2 of \citealt{CChang_2021}). While the Ray-Tracing algorithm is largely successful in predicting the presence and location of artifacts in the images, this algorithm is also limited \NEW{in predicting the amplitude of the ghost image} by the accuracy of the optical model and the external star catalogs used.

Recently, \citet{CChang_2021} demonstrated an alternative approach using a convolutional neural network \citep[CNN;][]{Lecun_1998} to classify DES images containing ghosts and scattered-light artifacts. 
CNNs constitute a class of deep neural networks that are inspired by the visual cortex and optimized for computer vision problems. Since their invention, CNNs have found numerous applications in the field of astronomy, including galaxy morphology prediction \citep[e.g.,][]{Dieleman2015,Cheng_TY:2021}, star-galaxy separation \citep[e.g.,][]{Kim2016}, identification of strongly lensed systems \citep[e.g.,][]{Lanusse2018,Davies2019,Bom2019,Huang:2020,Huang:2021}, classifying galaxy mergers \citep[e.g.,][]{Ciprijanovic2021}, and many other applications. 
The CNN developed by \citet{CChang_2021} was able to predict whether an image contained ghosts or scattered-light artifacts with high-accuracy ($\sim 96\%$ in the training set, $\sim 86\%$ in the test set), but did not identify the specific pixels of the image that were affected by the presence of artifacts. Since ghosts and scattered-light artifacts often affect a subregion of an image, flagging entire images rejects a significant amount of high-quality data.

In contrast to classic CNNs, object detection algorithms are designed to determine the location  of objects in an image (e.g., place bounding boxes around objects or mask exact pixels that belong to objects). In this work, we study the use of a deep learning-based object detection algorithm, namely a Mask Region-Based Convolutional Neural Network \citep[Mask R-CNN;][]{He_2017}, to predict the location of ghosts and scattered-light artifacts in astronomical survey images. Mask R-CNNs have recently been demonstrated as an accurate tool to detect, classify, and deblend astronomical sources (stars and galaxies) in images \citep{Burke_2019}.

Using 2000 manually annotated images, we train a Mask R-CNN model to identify artifacts in DES images. Comparing the results to those of the Ray-Tracing algorithm on ghost-containing images, we find that Mask R-CNN performs better in masking affected regions --- indicated by the value of the $F1$ score (a combination of precision and recall).
This demonstrates that deep learning-based object detection algorithms can be effective in helping to address a challenging problem in astronomical surveys without any {\it a priori} knowledge of the optical system used to generate the images. 



This paper is organized as follows. In Sec.~\ref{sec: Data}, we present the dataset, including the annotation process, used in this work. In Sec.~\ref{sec: Methods}, we describe the Mask R-CNN algorithm, implementation, and the training procedure. In Sec.~\ref{sec: Results} we present results from the Mask R-CNN model, including examples of predicted masks, custom and commonly used evaluation metrics, and we compare its performance to that of a conventional algorithm. We further summarize our results and their applications, and conclude in Sec.~\ref{sec: Summary_and_Conclusions}.
The code and data related to this work are publicly available at the GitHub page of this project: {\url{https://github.com/dtanoglidis/DeepGhostBusters}}.

\section{Data}
\label{sec: Data}

In this section, we describe the datasets used for training and evaluating the performance of the Mask R-CNN algorithm for detecting ghosts and scattered-light artifacts. We briefly describe the DES imaging data, our manual annotation procedure, the creation of masks, and the agreement between the human annotators who performed these tasks.

\subsection{Dark Energy Survey Data}
\label{sec: Images}

DES is an optical/near-infrared imaging survey that completed six years of observations in January 2019.
The DES data cover $\sim 5000$ deg$^2$ of the southern Galactic cap in five photometric filters, $grizY$, to a depth of $i \sim 24$ mag \citep{DES_DR2}. 
The observations were obtained with DECam, a 570-megapixel camera mounted on the 4m Blanco Telescope at the Cerro Tololo Inter-American Observatory (CTIO) in Chile \citep{Flaugher_2015}. The focal plane of DECam consists of 62 $2048\times 4096$-pixel red-sensitive scientific charge-coupled devices (CCDs), while its field-of-view covers 3 deg$^2$ with a central pixel scale of $0.263''$.

Our data come from the full six years of DES observations \citep{DES_DR2}. 
For the training, validation, and testing of the Mask R-CNN model, we use 2000 images that cover the full DECam focal plane and are known to contain ghosts and scattered-light artifacts. 
These are part of the positive sample used in \citet{CChang_2021} to train a CNN classifier to distinguish between images with and without ghosts.  This dataset was assembled by selecting images that the Ray-Tracing program identified as likely to contain ghosts, and subsequently visually inspecting them to correct for false detections. 

As described in \citet{CChang_2021}, the image data were down-sampled images of the full DECam focal plane. Images were produced with the STIFF program \citep{Bertin:2012}, assuming a power-law intensity transfer curve with index $\gamma = 2.2$. Minimum and maximum intensity values were set to the 0.005 and 0.98 percentiles of the pixel value distribution, respectively. The pixel values in each image were then normalized to a range whose minimum and maximum corresponded, respectively, to the first
quartile $Q_1(x)$ and third quartile $Q_3(x)$ of the full distribution in the image, by multiplying each pixel value, $x_i$, by a factor $s_i = \frac{x_i-Q_1(x)}{Q_3(x)-Q_1(x)}$.  
Focal plane images were originally derived as $800 \times 723$-pixel, 8-bit grayscale images in Portable Network Graphics format, which were then downsampled to $400 \times 400$ pixels for use with the Mask R-CNN.
The data from \citet{CChang_2021} are publicly available.\footnote{\url{https://des.ncsa.illinois.edu/releases/other/paper-data}}

\subsection{Annotation process}
\label{sec: Annotation_pr}

\begin{figure*}[!ht]
\centering
\subfigure[]{\includegraphics[width=0.44\textwidth]{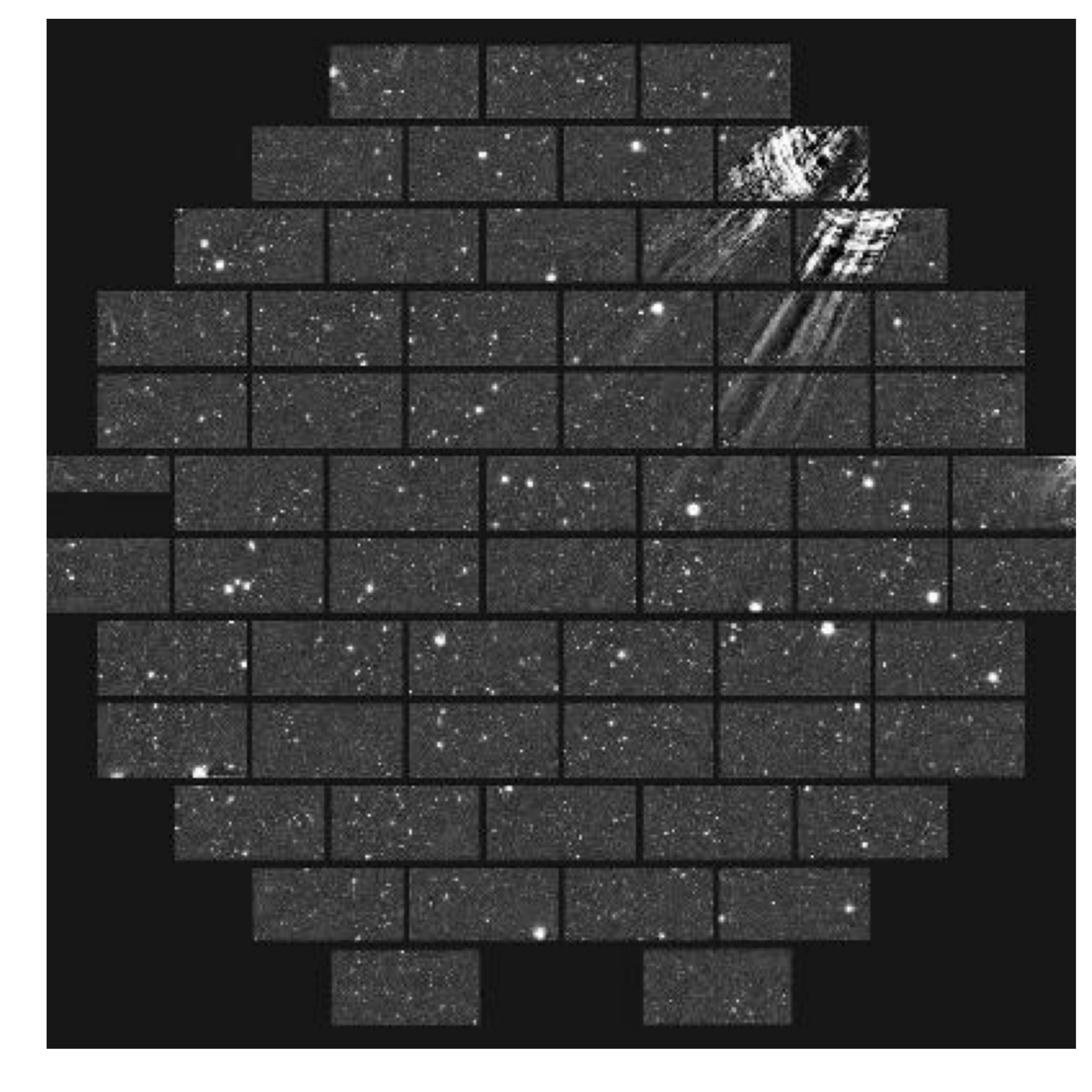}}
\hspace*{0.15cm}
\subfigure[]{\includegraphics[width=0.44\textwidth]{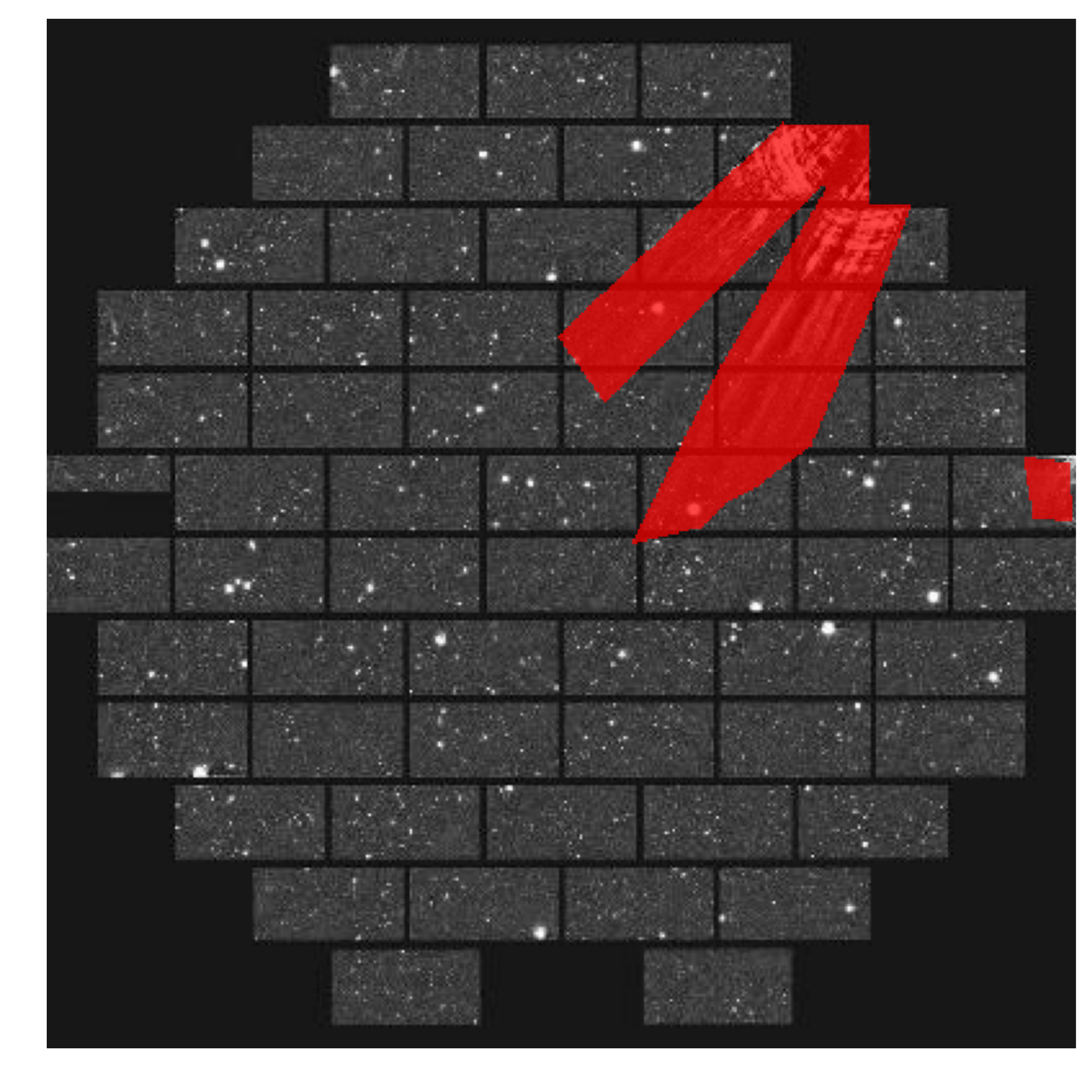}}
\subfigure[]{\includegraphics[width=0.44\textwidth]{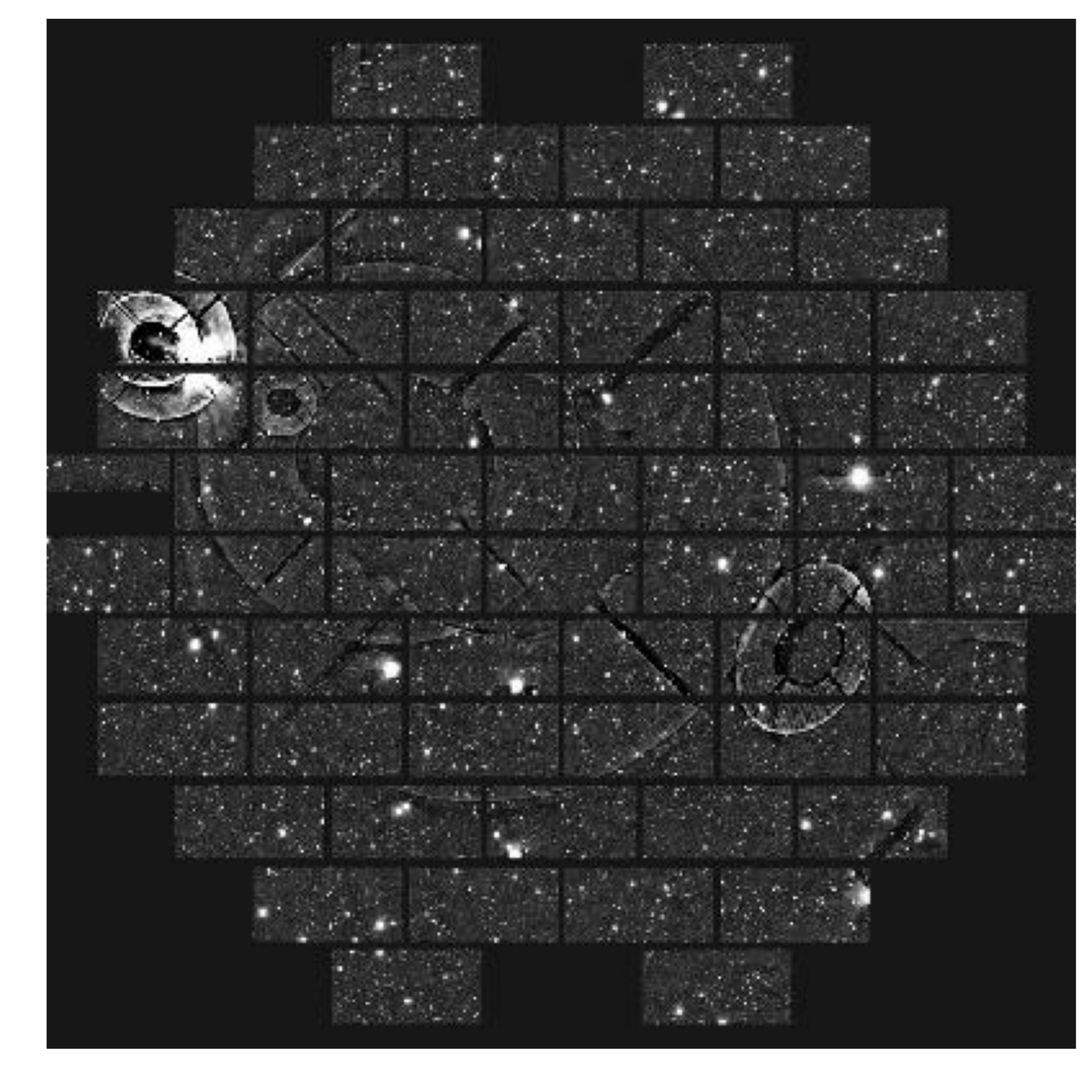}}
\hspace*{0.15cm}
\subfigure[]{\includegraphics[width=0.44\textwidth]{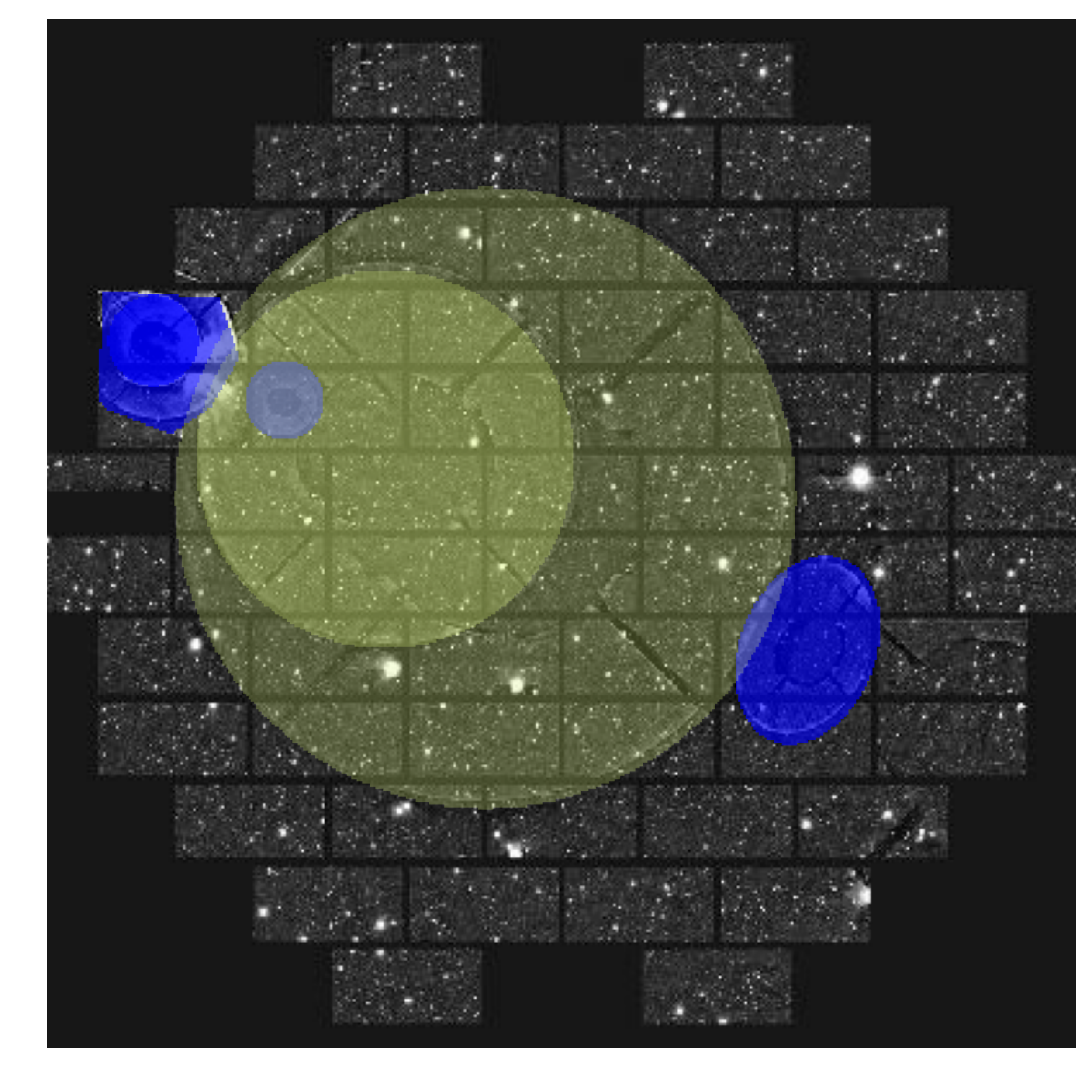}}
\vspace{-0.4cm}
\caption{Examples of full-focal-plane DECam images containing ghosts and scattered-light artifacts. The corresponding ``ground truth'' masks (right) were manually annotated. There are three categories of ghosting artifacts: image (a) contains a scattered-light artifact classified as `Rays'; image (b) shows the masks for the `Rays' in red; image (c) contains both `Bright' and `Faint' ghosts, and the corresponding masks in blue and yellow, respectively, are shown in image (d).}
\label{fig: Images_and_masks}
\end{figure*}

Training the Mask R-CNN algorithm requires both images and ground-truth segmentation masks identifying objects of interest in each image. 
To create these masks, we used the VGG Image Annotator (VIA; \citet{dutta2019vgg})\footnote{\url{https://www.robots.ox.ac.uk/~vgg/software/via/}}, a simple manual annotation software for images, audio, and video. We split the 2000 images into batches of 100 images, and we randomly assigned each batch to one of eight authors for annotation.\footnote{Note that not every author annotated the same number of images; six of us annotated 200 images and two of us annotated 400 images.} 

During manual annotation, we categorized the ghosting and scattered-light artifacts into three distinct morphological categories:
\begin{enumerate}
    \item `Rays': These are scattered-light artifacts originating from the light of off-axis stars scattering off of the DECam filter changer \citep{Kent:2013}. They emanate from one of the edges of the image and span several CCDs. This is the most distinct artifact category and is not commonly confused with either of the other two categories.
    \item `Bright': These are high-surface-brightness ghosting artifacts that come from multiple reflections off the DECam focal plane and the C4 or C5 lenses \citep{Kent:2013}. They are usually relatively small in size and circular or elliptical in shape. They have more distinct borders and are considerably brighter compared to the following category. 
    \item `Faint': These are lower-surface-brightness ghosting artifacts that come from multiple reflections between the focal plane and the C3 lens or filter, or internal reflections off of the faces of the C3, C4, and C5 lenses \citep{Kent:2013}. They are circular or elliptical in shape and are usually larger in size and significantly fainter than `Bright' ghosts. 
\end{enumerate}
In Fig.~\ref{fig: Images_and_masks}, we present two examples of DECam images that contain ghosts and scattered-light artifacts, along with the annotated ground truth masks. 
We trained the Mask R-CNN for these three distinct categories due to their significant morphological difference. 

\begin{figure}[t]
\centering
\includegraphics[width=1.0\columnwidth]{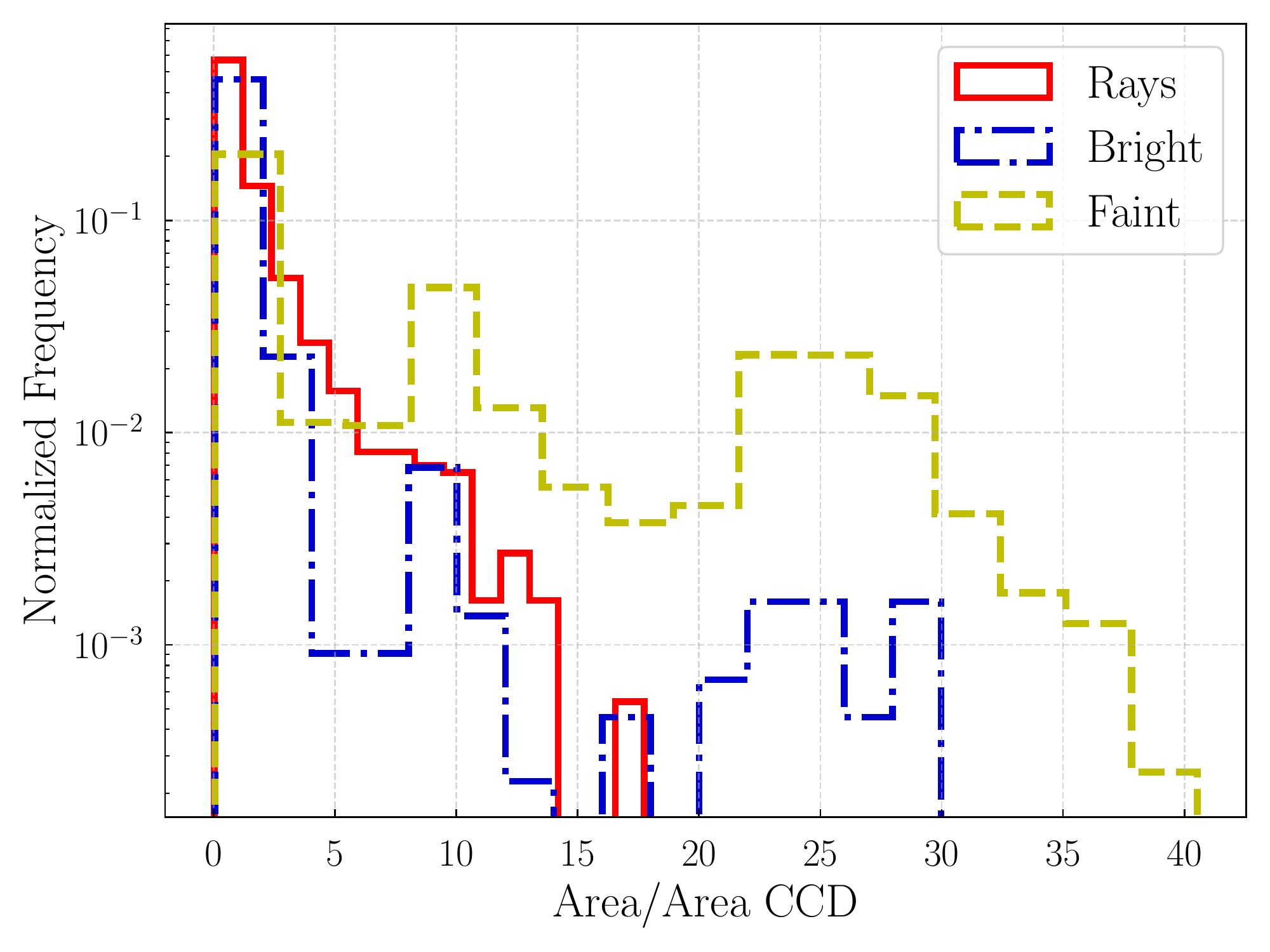}
\caption{Histograms of the distribution in size (area) of the three artifact types presented in this work. The areas are quoted as multiples of the area of a single CCD.}
\label{Fig: Ghost_Areas}
\end{figure}

In total, our dataset contains 1566 `Rays', 2197 `Bright', and 2949 `Faint' artifact instances. In Fig.~\ref{Fig: Ghost_Areas}, we present the distribution in size (area) of these three ghost categories. The area of each ghosting artifact is presented as a fraction of the area of a single DECam CCD (area of artifacts in pixel over area of a CCD in pixels). Most `Rays' have an area that covers fewer than 10 CCDs. `Bright' ghosts are also relatively small in size, with a few spanning more than a couple of CCDs. On the other hand, `Faint` ghosts are large in size, with a significant fraction of them covering an area of 20--30 CCDs. Many images contain multiple ghosts or scattered-light artifacts.

\begin{figure*}
\centering
\subfigure[]{\includegraphics[width=0.48\textwidth]{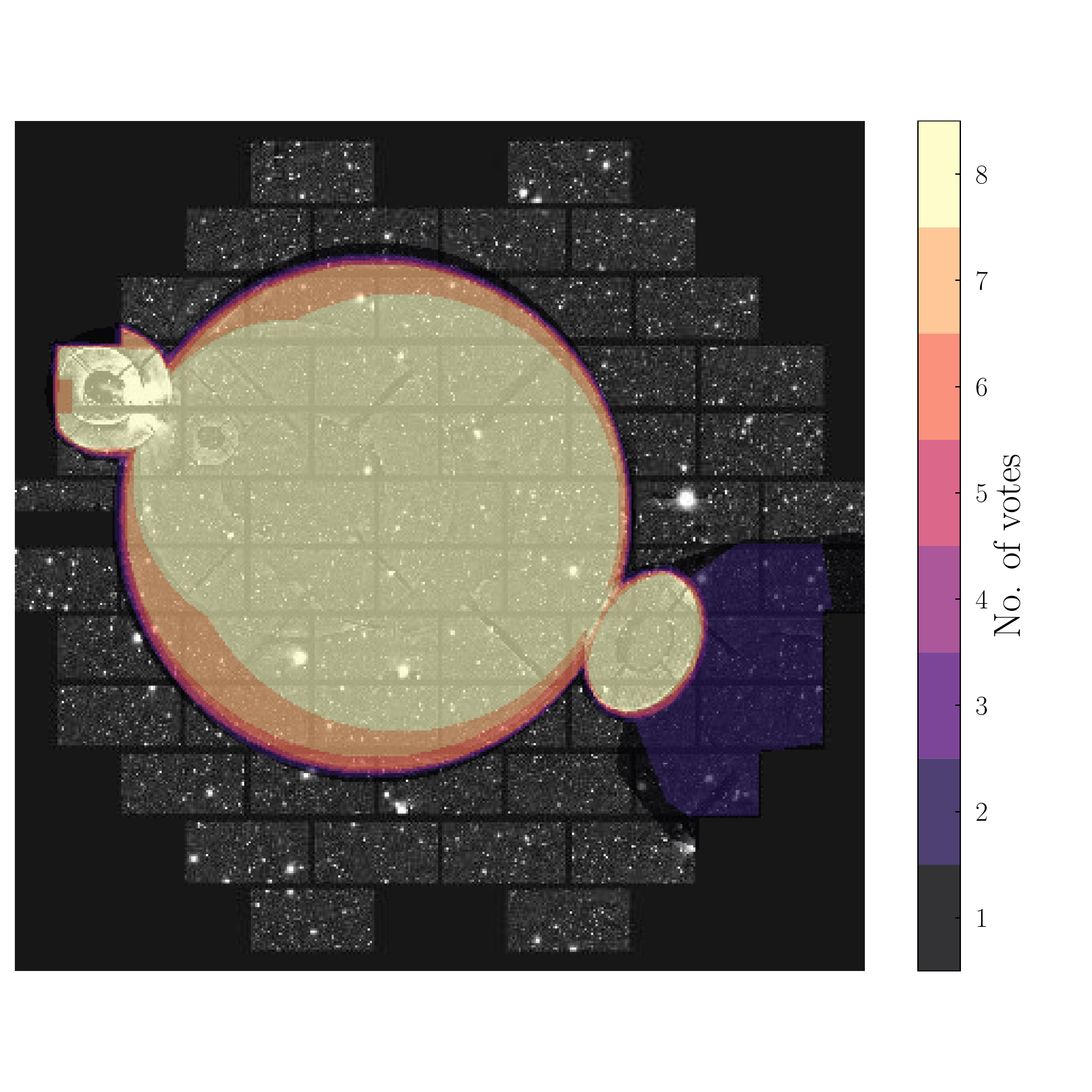}}
\hspace*{0.1cm}
\subfigure[]{\includegraphics[width=0.48\textwidth]{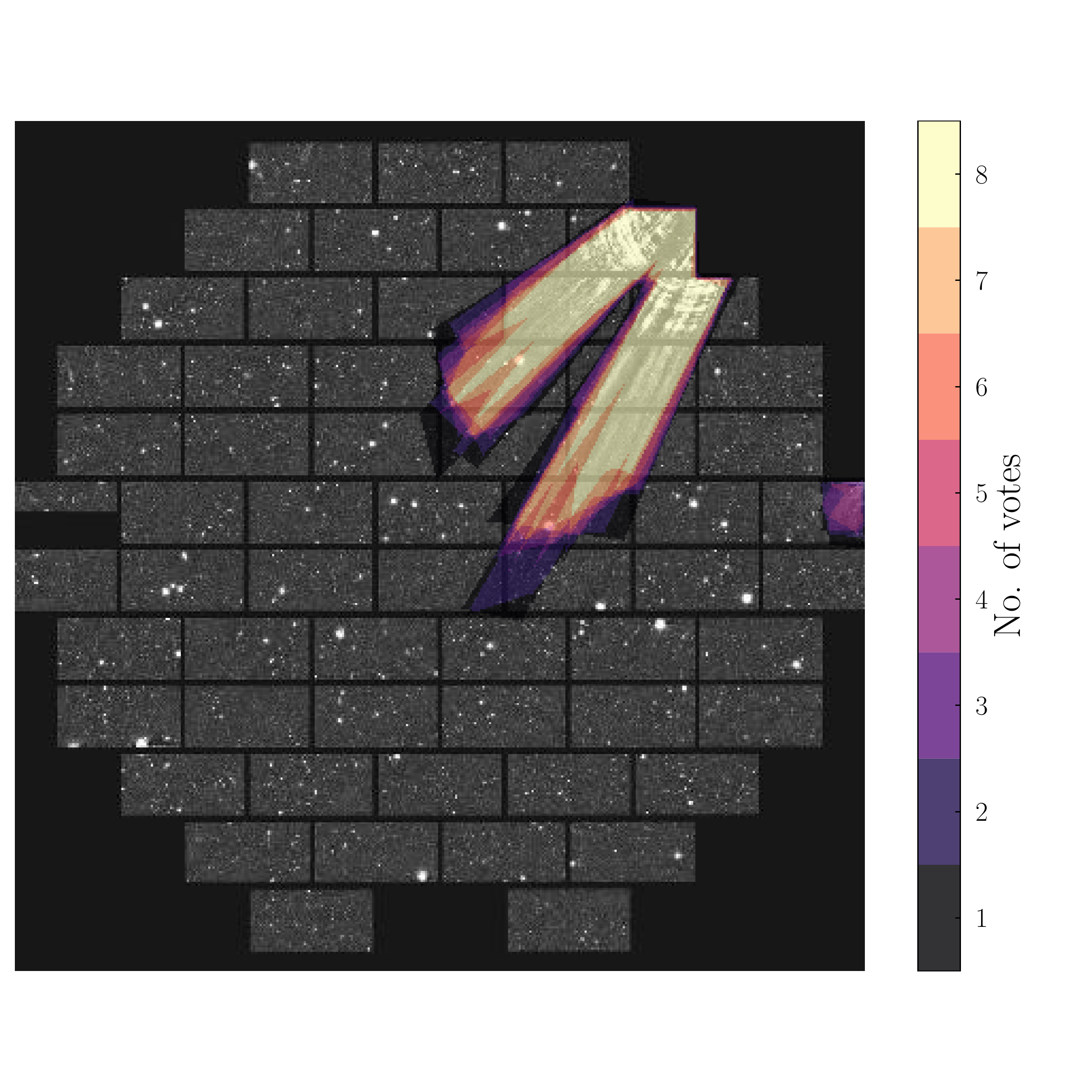}}
\vspace{-0.4cm}
\caption{Masks created by the eight different annotators (overlaid on top of each other) for the same two images presented in Fig.~\ref{fig: Images_and_masks}. The colors indicate the number of annotators that have labeled a given pixel as containing a ghost, from dark purple (one annotator) to light yellow (all the eight annotators).}
\label{fig: Masks_Annotators}
\end{figure*}

We note that the ghosting and scattered-light artifacts do not always have clear boundaries (especially those of type `Rays') and that the distinction between `Bright' and `Faint' ghosts is not always well defined. For that reason we expect some disagreement between the human annotators in the extent and shape of the ground truth masks and in the assigned labels. 

In Fig.~\ref{fig: Masks_Annotators}, we overlay the masks generated by all eight annotators for the same two DECam images presented in Fig.~\ref{fig: Images_and_masks}. The colors correspond to the number of annotators that have labeled the region as containing an artifact; dark purple corresponds to fewer votes, while light yellow corresponds to more votes. We do not distinguish between the different artifact types in this image.

The right panel of Fig.~\ref{fig: Masks_Annotators} shows a significant variation in the masks created by the different annotators for the `Rays'. The left panel shows generally good agreement between the different annotators for the most prominent ghosts in the image; however, there is a large area on the right of the image that is labeled by only two annotators.
We discuss the agreement between the human annotators in more detail in \ref{sec: Agreement_between_annot}. 
In Section~\ref{sec: Results}, we demonstrate that the Mask R-CNN is able to out-perform conventional algorithms even in the presence of the label noise introduced by disagreements in the existence, mask region, and classification of artifacts by individual annotators.
Reduction in label noise from more uniform annotation could improve the performance of the algorithm in the future.

\section{Methods}
\label{sec: Methods}

We use Mask R-CNN \citep{He_2017}, a popular, state-of-the art instance segmentation algorithm, to detect and mask ghost and scattered-light artifacts.

Mask R-CNN is a powerful and complex algorithm, the latest in a series of object detection models, collectively known as the R-CNN family.\footnote{Mask R-CNN is the latest in the R-CNN family for 2D object detection. Mesh R-CNN \citep{Gkioxari_2019} is a more recent addition to the family, and it is able to predict 3D shapes of the detected objects.} It builds upon many deep learning and computer vision techniques; we refer the reader to \citet{Weng_2017} for a detailed description of the R-CNN family.

Instance segmentation \citep[e.g., for a review,][]{Hafiz:2020} combines the functions of object detection and image segmentation algorithms. Object detection \citep[e.g., for a review,][]{Zhao:2018} is an active area of research in computer vision, with the goal of developing algorithms that can find the positions of objects within an image. Semantic segmentation \citep[e.g., for a review,][]{Minaee:2020} on the other hand refers to the problem of pixel-level classification of different parts of an image into pre-defined categories. 
Instance segmentation is used to simultaneously detect objects in an image and to create a segmentation mask for each object.


\begin{figure}[ht]
\vspace{-0.4cm}
\centering
\includegraphics[width=1.0\columnwidth]{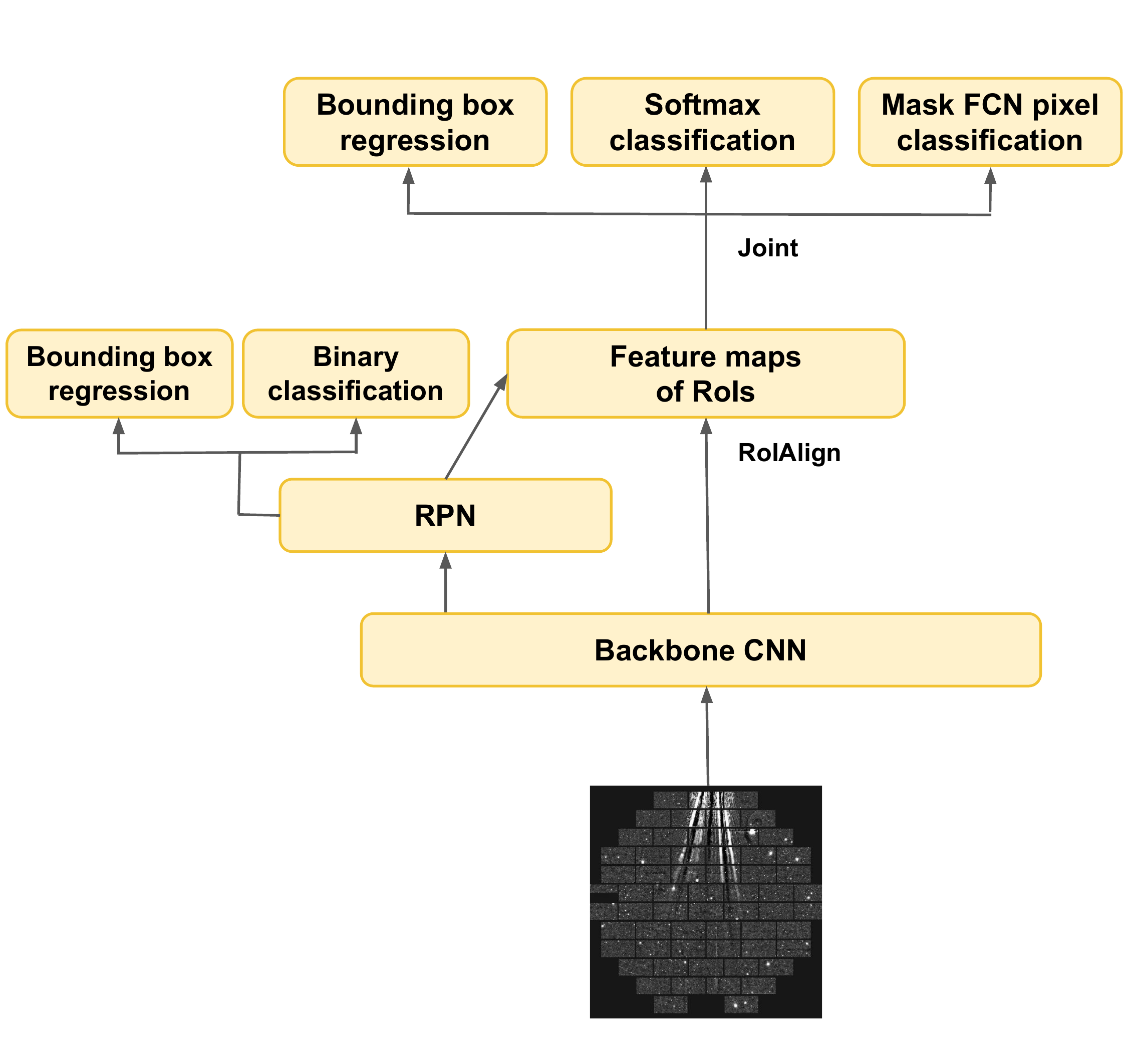}
\caption{High-level schematic overview of the Mask R-CNN model. Figure adapted from \citet{Weng_2017}.}
\label{Fig: Mask_R_CNN}
\end{figure}


A schematic description of the Mask R-CNN workflow is presented in Fig.~\ref{Fig: Mask_R_CNN}. In the first stage of the model, the input images are fed into a pre-trained deep CNN --- such as VGG \citep{VGG_Simonyan} or ResNet \citep{He_2015} --- also called the \textit{backbone} network. The last, fully connected, classification layers of this network have been removed, and thus its output is a feature map. This feature map\footnote{In practice, most Mask R-CNN implementations -- like the one we are using in this work -- use a Feature Pyramid Network \citep[FPN;][]{Lin_Tsung_2016} on top of the backbone. The FPN combines low-level features extracted from the initial stages of the backbone CNN with the high-level feature map output of the last layer. This improves the overall accuracy of the model, since it better represents object at multiple scales.} is  passed into the Region Proposal Network (RPN) to produce a limited number of Regions of Interest (RoIs) to be passed to the main network -- i.e., candidate regions that are most likely to contain an object.

The RPN is a simple CNN that uses a sliding window to produce a number of \textit{anchor boxes} -- boxes of different scales and aspect ratios -- at each position. When training the RPN network, two problems are considered --- classification and regression. For classification, the algorithm considers the possibility that there is an object (without considering the particular class) that fits inside an anchor box. For regression, the best anchor box coordinates are predicted. The anchor boxes with the highest object-containing probability scores are passed as RoIs in the next step. 
The loss of the RPN network is composed of a binary classification loss, $L_{\mbox{\scriptsize{RPN,cls}}}$, and a bounding box regression loss, $L_{\mbox{\scriptsize{RNP,bbox}}}$, such that $L_{\mbox{\scriptsize{RPN}}}=L_{\mbox{\scriptsize{RPN,cls}}}+L_{\mbox{\scriptsize{RNP,bbox}}}$.

Each of the proposed RoIs has a different size. However, the fully connected networks used for prediction require inputs of the same size. For that reason, the RoIAlign method is used to perform a bilinear interpolation on the feature maps within the area of each RoI and output the interpolated values within a grid of specific size, giving fixed-size feature maps of the candidate regions.

Finally, these reshaped regions are passed to the last part of the Mask R-CNN that performs three tasks in parallel. A softmax classifier learns to predict the class of the object within the RoI; the output is one of the $K+1$ classes, where $K$ are the different possible object types ($L_{\mbox{\scriptsize{cls}}}$ loss), plus one background class. A regressor learns the best bounding box coordinates ($L_{\mbox{\scriptsize{bbox}}}$ loss). Finally, the regions pass through a Fully Convolutional Network (FCN) that performs semantic segmentation ($L_{\mbox{\scriptsize{mask}}}$ loss), i.e. a per-pixel classification, that creates the masks. The total loss of this Mask R-CNN part is thus $L_{\mbox{\scriptsize{tot}}}=L_{\mbox{\scriptsize{cls}}}+L_{\mbox{\scriptsize{bbox}}}+L_{\mbox{\scriptsize{mask}}}$.
 
The {\it DeepGhostBusters} algorithm is the Mask R-CNN implementation by \citet{matterport_maskrcnn_2017}, trained on our manually annotated dataset of ghosting and scattered-light artifacts.  
This code is written in Python using the high-level \texttt{Keras}\footnote{\url{https://keras.io/}} library using a \texttt{TensorFlow}\footnote{\url{https://www.tensorflow.org/}} backend. 
We use the default 101-layer deep residual network (ResNet-101; \citealt{He_2015}) as the backbone convolutional neural network architecture.

\begin{figure}[!ht]
\centering
\includegraphics[width=1.0\columnwidth]{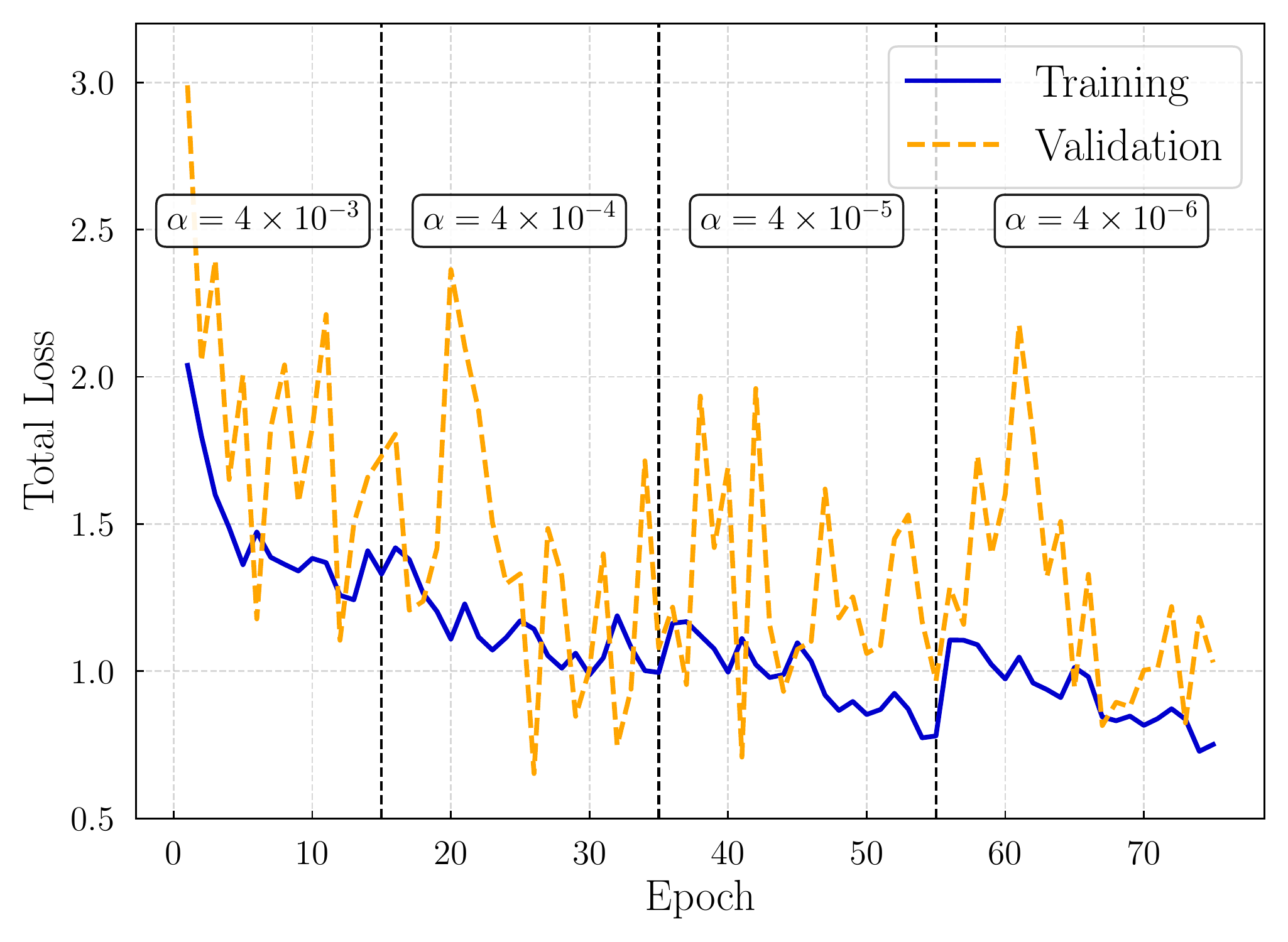}
\caption{Total loss of the Mask R-CNN model as function of the training epoch. The training is performed using a progressively smaller learning rate, $\alpha$.}
\label{Fig: Total_Loss}
\end{figure}

Before training, we randomly split the full dataset of 2000 images into a training set (1400 images), a validation set (300 images), and a test set (300 images). The annotation process was performed before this random split. Such a random split is generally important in machine learning problems for these three sets to be representative of the general population, but it becomes even more important here because different human annotators have different annotation styles. This could create significant systematic differences between the ground truth masks in the datasets if not properly randomized.

In computer vision problems where only a small training set is available, it is common to use \textit{transfer learning} to improve results (for recent reviews, see \citealt{Wang:2018} and \citealt{Zhuang:2019}). Transfer learning is a process where the weights of a network that has already been trained for one detection task are used for a different, but related, task, usually with some further training. This speeds up the training process, reduces overfitting, and produces more accurate results. Here, we initialize the learning procedure (i.e., use transfer learning) using the weights learned from training Mask R-CNN on the Microsoft Common Objects in Context (MS COCO) dataset\footnote{\url{https://cocodataset.org/\#home}} \citep{MS_COCO}, which consists of $\sim 330$k images ($\sim 2.5$M object instances) of 91 classes of common or everyday objects.

To reduce overfitting, we employ data augmentation \citep[e.g.,][]{Shorten:2019}, by performing geometric transformations on the images and the masks. Specifically, we randomly apply zero to three of the following transformations:
\begin{itemize}
    \item Rotation of the image and the masks by $270$ degrees.
    \item Left-right mirroring/flip of the images and masks.
    \item Up-down mirroring/flip of the images and masks.
\end{itemize}

We re-train our model using stochastic gradient descent to update the model parameters. Similarly to what was proposed in \citet{Burke_2019}, the training is performed in different stages with progressively smaller learning rates, $\alpha$, at each stage. This allows for a deeper learning and finer tuning of the weights, while minimizing the risk of overfitting.

Specifically, in the first stage (15 epochs), we re-train the top layers only and use a learning rate of $\alpha=4\times 10^{-3}$. Then, we train all the layers with decreasing learning rates: 20 epochs at $\alpha=4\times10^{-4}$, 20 epochs at $\alpha=4\times10^{-5}$, and 20 epochs at $\alpha=4\times10^{-6}$. In total, we trained the model for 75 epochs, after which overfitting occurs. In all stages (training, validation, test) we ignore detections with less than $80\%$ confidence (\texttt{DETECTION\_MIN\_CONFIDENCE = 0.8}).
We utilized the 25 GB high-RAM Nvidia P100 GPUs available through the Google Colaboratory (Pro version). The training took $\sim 4$ hours to complete. The inference time is $\sim 0.34$s per image to predict.

In Fig.~\ref{Fig: Total_Loss}, we present the total loss as a function of the training epoch for both training and validation sets. In \ref{sec: Losses}, we show the training history for the individual components of the total loss.

\section{Results}
\label{sec: Results}

We use an independent DECam test set to evaluate the performance of the {\it DeepGhostBusters} Mask R-CNN in detecting and masking ghost and scattered-light artifacts. We use both custom metrics appropriate for the problem at hand and metrics commonly used in the object detection literature. 
We also compare the performance of \textit{DeepGhostBusters} with  the conventional Ray-Tracing algorithm. 
Finally, we test the classification performance of \textit{DeepGhostBusters} when it is presented with a dataset that also contains images that lack any ghosts or scattered-light artifacts.

\subsection{Example Performance}
\label{sec: examples}

\begin{figure*}[!ht]
\centering
\subfigure[]{\includegraphics[width=0.44\textwidth]{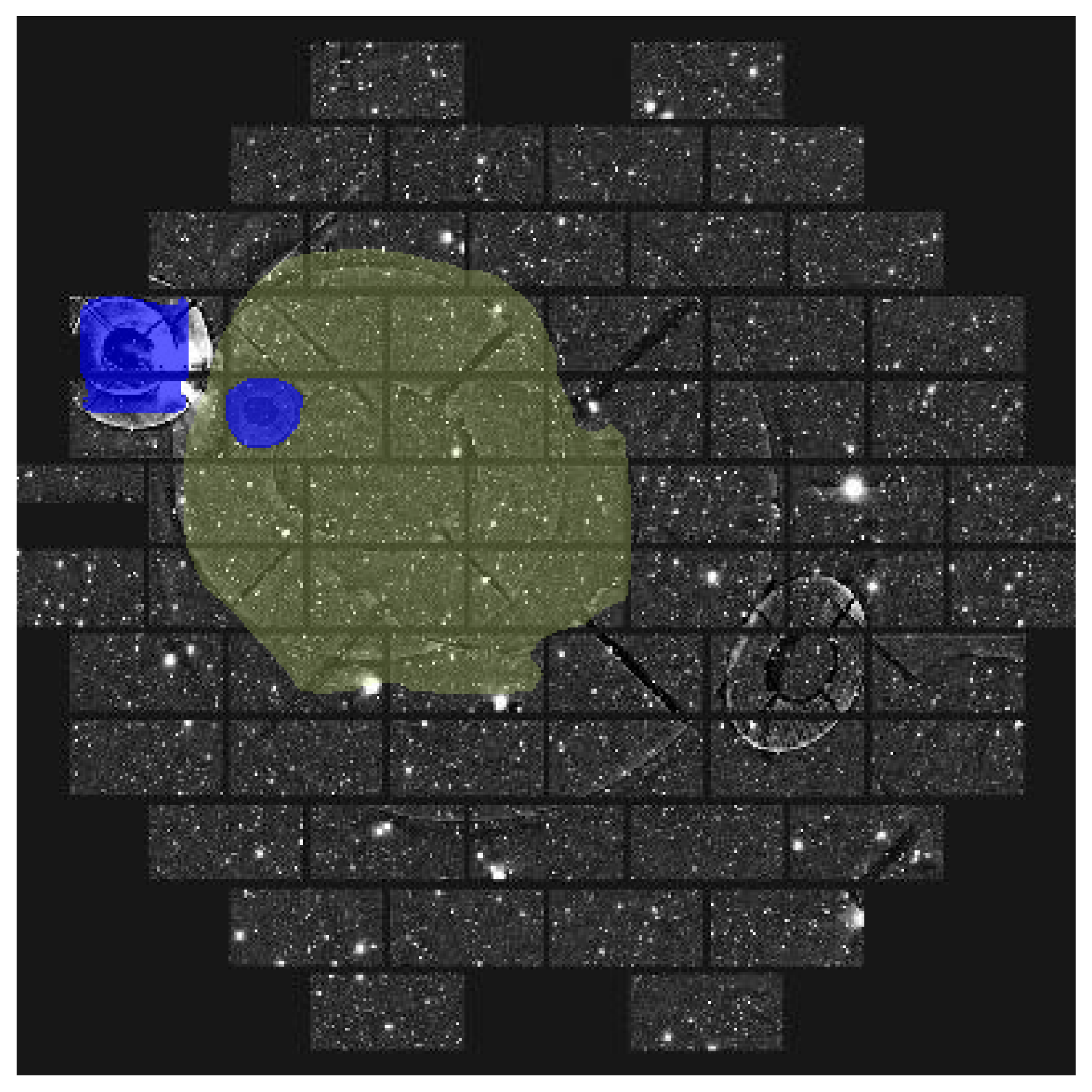}}
\hspace*{0.15cm}
\subfigure[]{\includegraphics[width=0.44\textwidth]{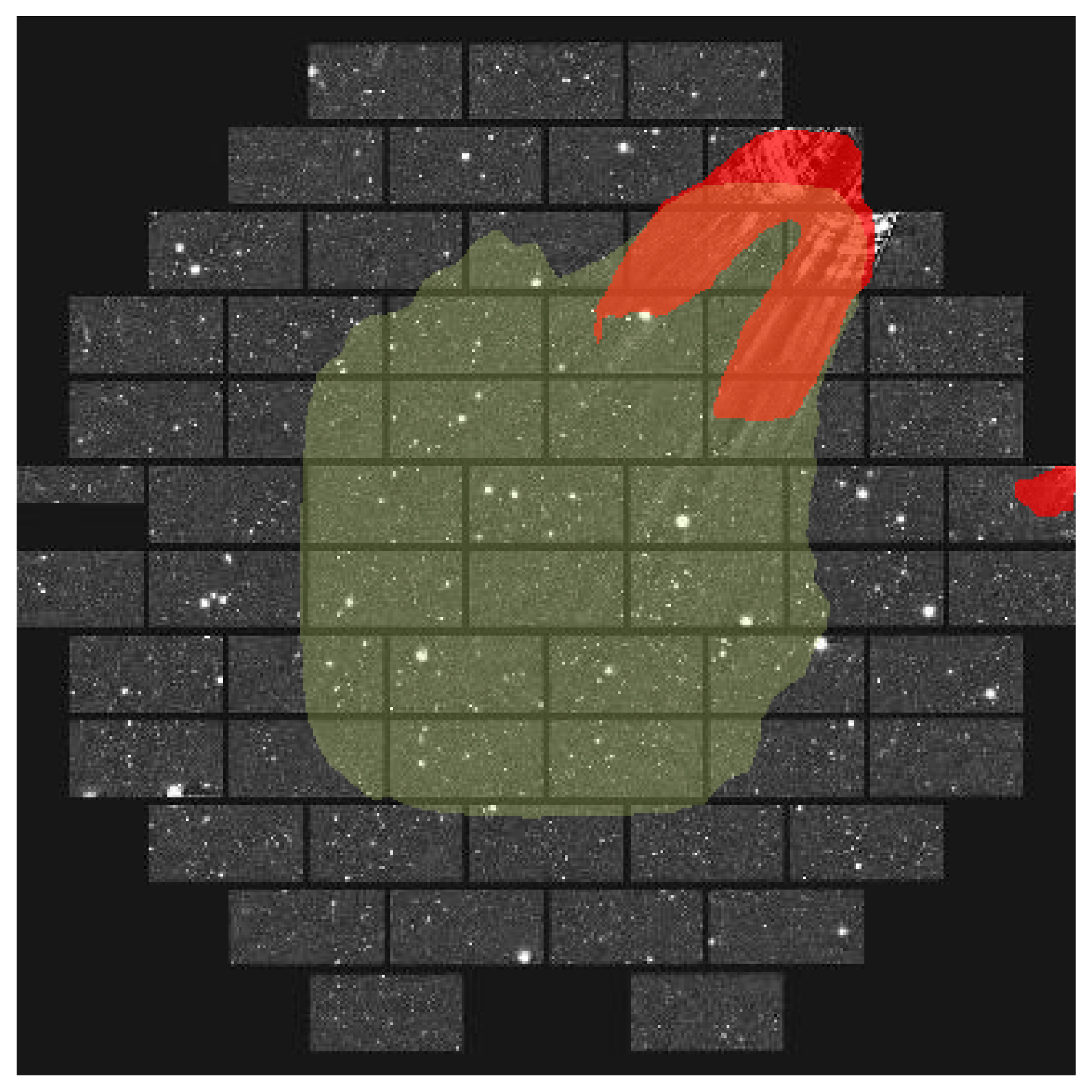}}
\subfigure[]{\includegraphics[width=0.44\textwidth]{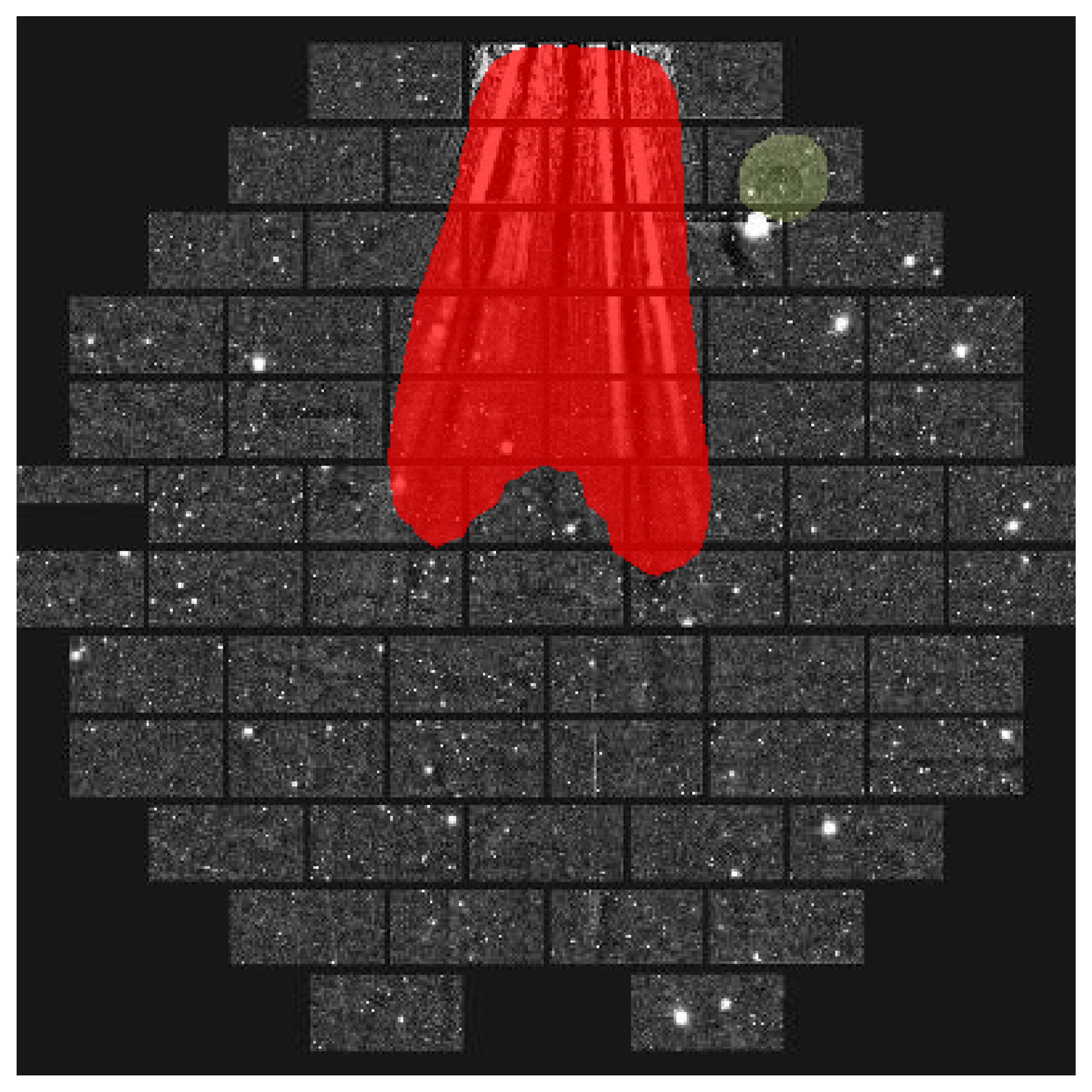}}
\hspace*{0.15cm}
\subfigure[]{\includegraphics[width=0.44\textwidth]{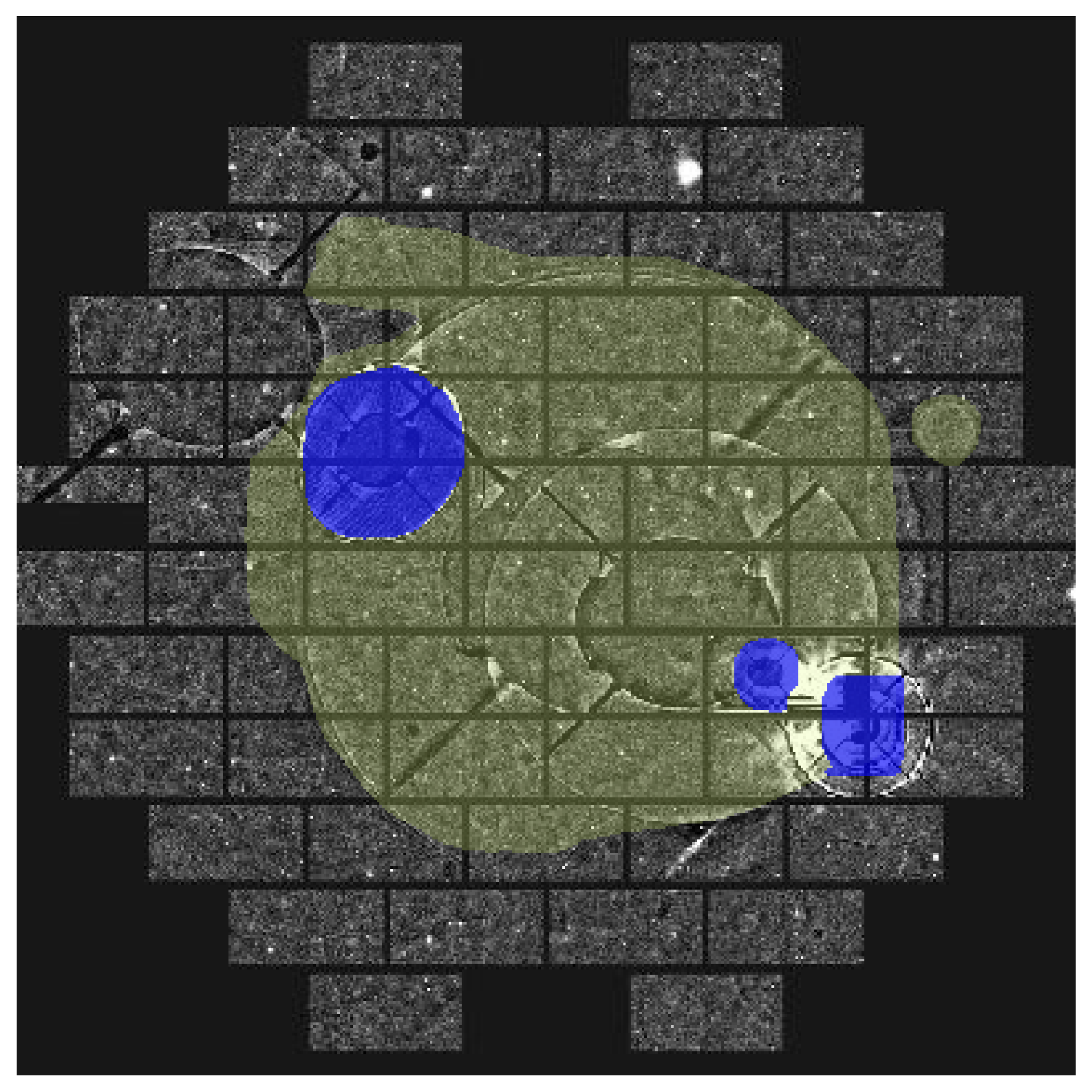}}
\vspace{-0.4cm}
\caption{Predicted masks on four example images that contain the three distinct artifact types --- scattered-light `Rays' (red), `Bright' ghosts (blue), and `Faint' ghosts (yellow). The top panels correspond to the images presented in Fig.~\ref{fig: Images_and_masks}.}
\label{fig: Prediction_Examples}
\end{figure*}

We first present the mask and class predictions of the {\it DeepGhostBusters} Mask R-CNN model on four example images (Fig.~\ref{fig: Prediction_Examples}). The two top panels, (a) and (b), correspond to the same images whose ground truth masks were presented in Fig.~\ref{fig: Images_and_masks}. As in Fig.~\ref{fig: Images_and_masks}, the different colors represent the different ghosting artifact types: red for `Rays', blue for `Bright', and yellow for `Faint'.

These examples demonstrate both the successes and failures of our model. For example, in panel (a) the model has successfully masked most of the central `Faint' ghost, but it has also missed a significant part of its periphery, as well as the prominent ghost on the right of the image. 
Furthermore, although it has successfully deblended and separately masked the small `Bright' ghost that is superimposed on the larger `Faint' one, it has only partially masked the one on the left.
Panel (b) presents a characteristic example of a false positive detection: predicting a mask for a `Faint' ghost that is not there.  The Mask R-CNN has predicted a mask that successfully covers most of the prominent `Rays'-type artifact; it is also able to detect the smaller `Rays' on the right. However, it has also erroneously masked a large central region (containing the edges of the rays) as a `Faint' ghost.
Panels (c) and (d) present mostly successful detections, although with some false negatives, as the undetected `Faint' ghost on the top-left corner of panel (d). 
We next formally quantify and evaluate the performance of the Mask R-CNN model and compare it with that of the conventional Ray-Tracing algorithm.

\subsection{CCD-based metrics}
\label{sec: CCD-based}

The DECam focal plane consists of 62 science CCDs. The conventional Ray-Tracing algorithm used by DES flags affected focal plane images on a CCD-by-CCD basis --- i.e., if a CCD contains a ghost or scattered-light artifact, the entire CCD is removed from processing. 
To compare the performance of the Mask R-CNN to the conventional algorithm, we develop metrics that are based on whether a CCD contains a ghost or scattered-light artifact. 

The resulting metrics depend on the size of individual artifacts. This is important for the problem at hand: for example, we care how well the algorithm can mask a larger ghost compared to a smaller one. At the same time, given the challenges of this problem (e.g., overlapping sources and borders that are not always well defined), assessing the performance at the CCD-level can be more robust than comparisons at the more granular pixel level.

We consider each image as a 1D array of length 62 with entries 0 and 1, where 0 corresponds to CCDs that do not contain a ghost, and 1 corresponds to those that do contain a ghost. For a batch of $M$ images containing $N = 62\times M$ CCDs, we define the number of true positives ($N^{TP}$), true negatives ($N^{TN}$), false positives ($N^{FP}$), and false negatives ($N^{FN}$). Then, we define the CCD-based precision (purity) and recall (completeness) as:

\begin{equation}
\label{eqn:precision}
\mbox{Precision}_{\scriptsize{CCD}} = \frac{N^{TP}}{N^{TP}+N^{FP}},
\end{equation}

\begin{equation}
\label{eqn:recall}
\mbox{Recall}_{\scriptsize{CCD}} = \frac{N^{TP}}{N^{TP}+N^{FN}}.
\end{equation}

\noindent \NEW{Based on the science case of interest, one may want to maximize either the precision or the recall. For example, for systematic studies of low-surface-brightness galaxies, high recall for ghosts and scattered-light artifacts may be preferred at the expense of some loss in precision. }

\NEW{One approach to assessing the trade-off between precision and recall is to define the $F1$ score, which is the harmonic mean of the precision and recall,}
\begin{equation}
\label{eqn:F1_score}
F1_{\scriptsize{CCD}} = 2 \left ( \frac{\mbox{Precision}_{\scriptsize{CCD}}\cdot \mbox{Recall}_{\scriptsize{CCD}}}{\mbox{Precision}_{\scriptsize{CCD}}+\mbox{Recall}_{\scriptsize{CCD}}} \right)
\end{equation}
Note that we can use the above definitions for each type of artifact individually or for all artifact types combined. 

\begin{figure*}[!ht]
\centering
\subfigure[]{\includegraphics[width=0.48\textwidth]{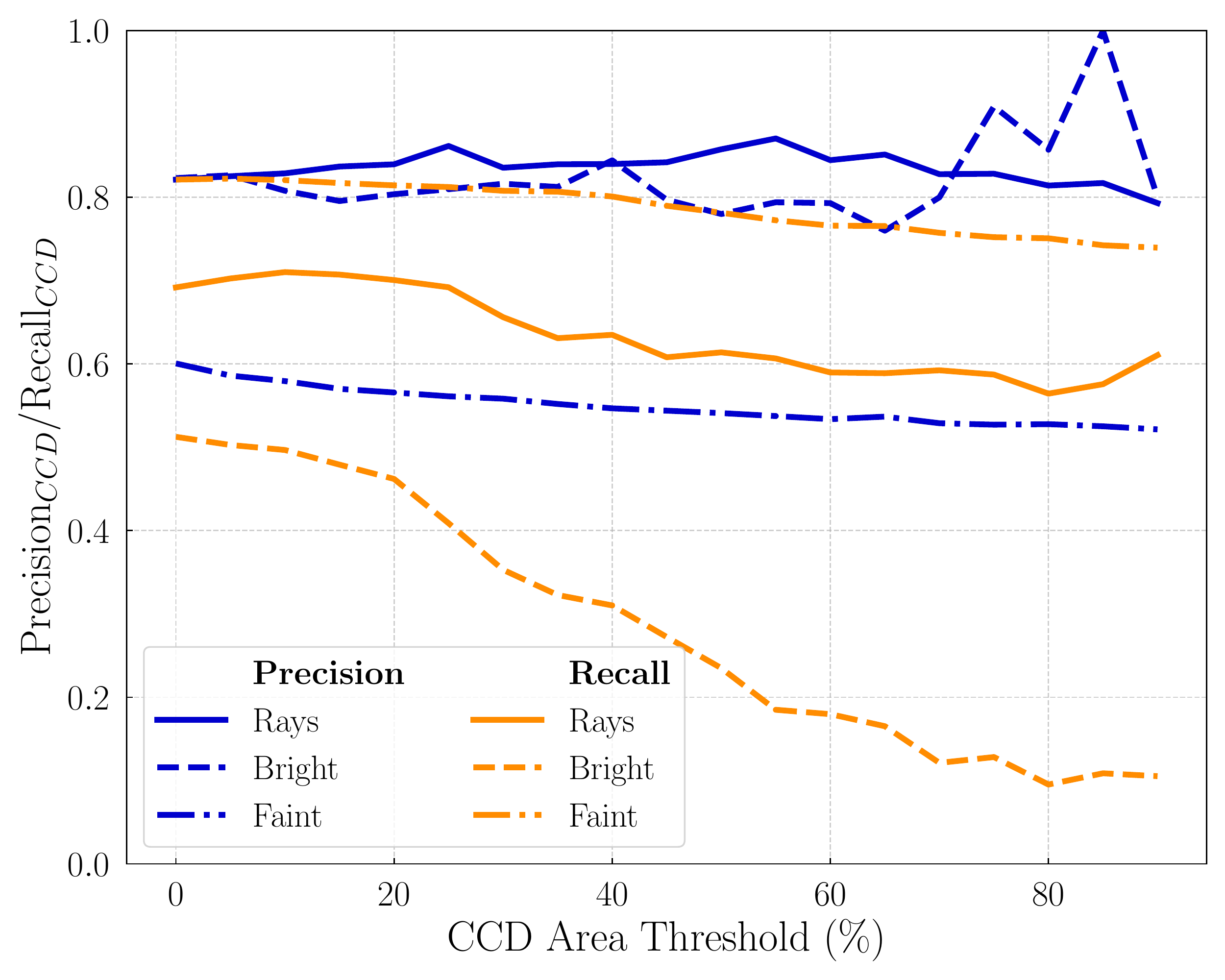}}
\hspace*{0.2cm}
\subfigure[]{\includegraphics[width=0.48\textwidth]{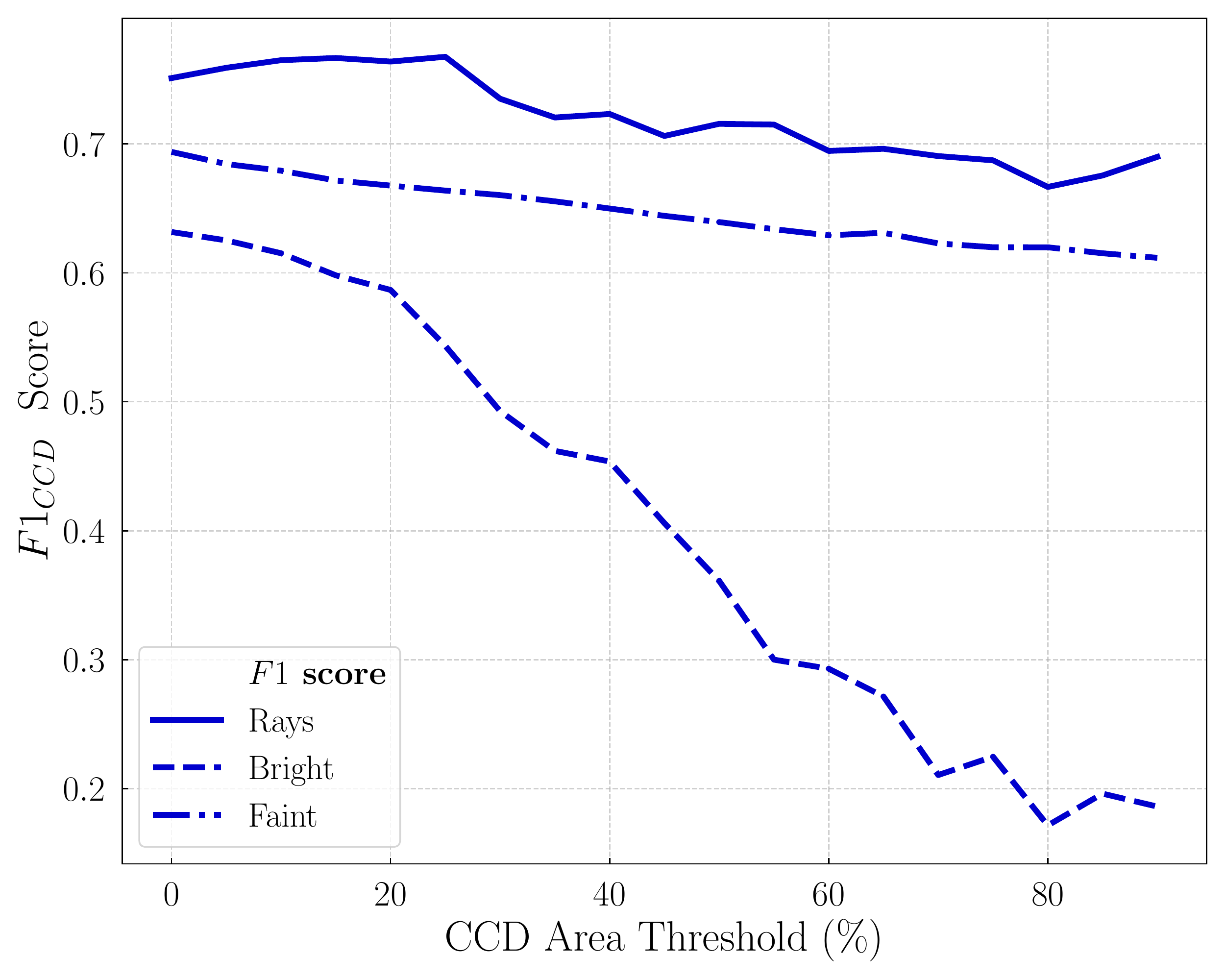}}
\vspace{-0.4cm}
\caption{CCD-based (a) precision (blue) and recall (orange), and (b) $F1$ score as a function of the CCD area threshold (see main text) from the Mask R-CNN model and for the three ghosting artifact categories (`Rays', `Bright', and `Faint').}
\label{Fig: Individual_metrics}
\end{figure*}

The above metrics are based on the notion of a binary classification of CCDs as affected by ghosts or scattered-light artifacts. 
In reality, the ghosts and scattered-light artifacts will only cover some fraction of the CCD area.
Thus, we define a threshold for the fraction of the CCD area that must be covered for the CCD to be classified as affected. In \ref{sec: masking} we present examples of masked CCDs for two different area thresholds.
Here, we study how the performance metrics change as a function of that threshold.


\begin{figure*}[!ht]
\centering
\subfigure[]{\includegraphics[width=0.48\textwidth]{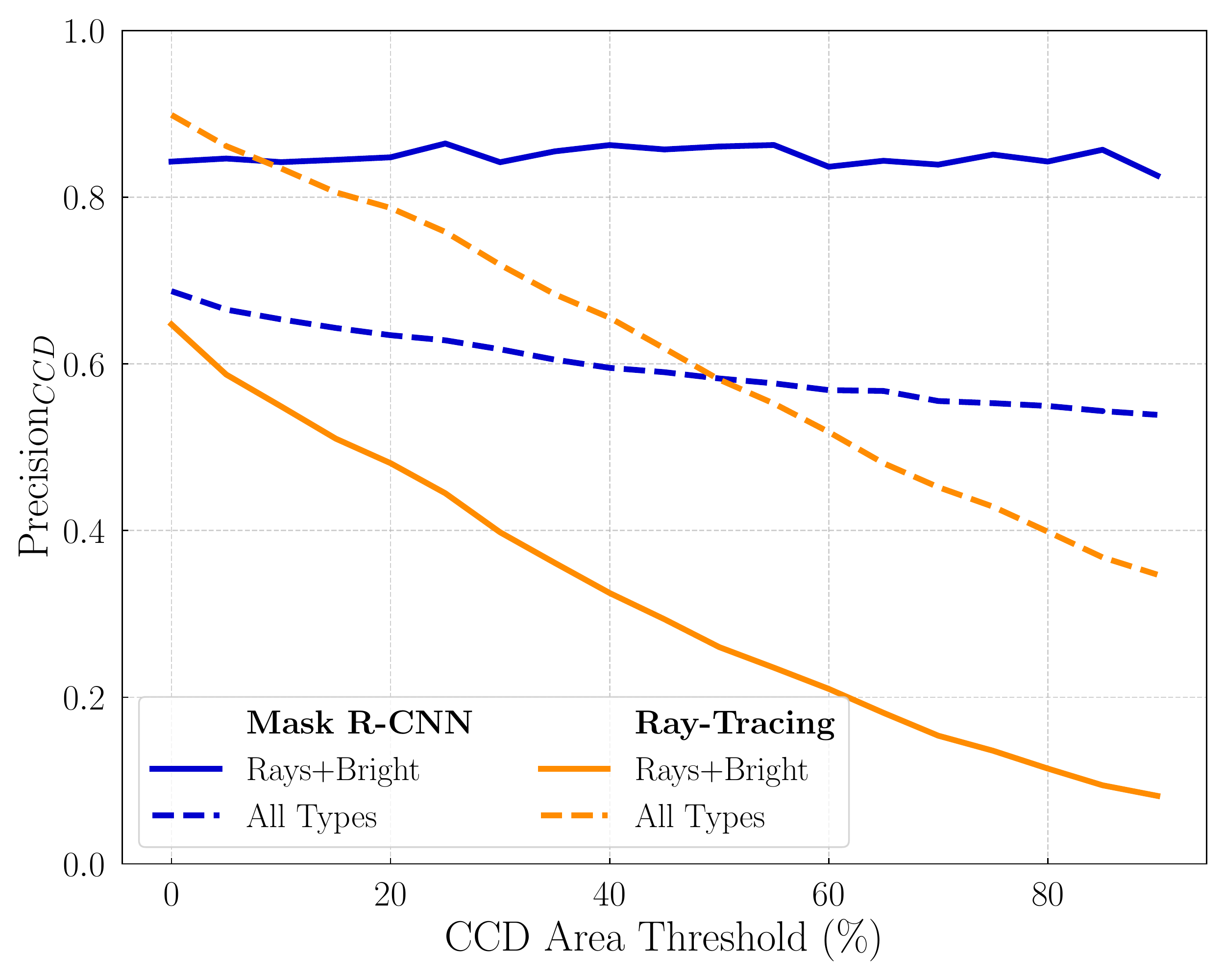}}
\hspace*{0.2cm}
\subfigure[]{\includegraphics[width=0.48\textwidth]{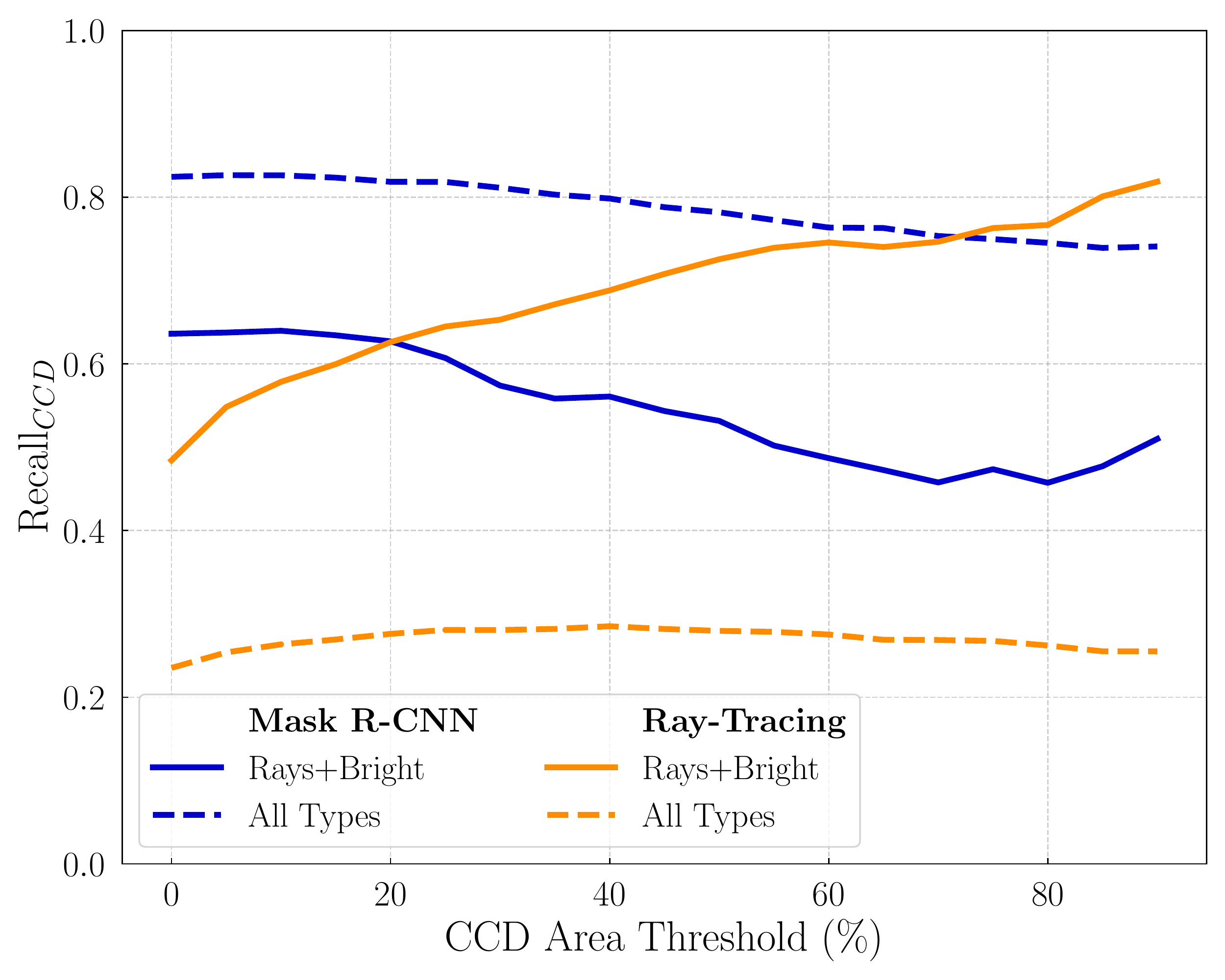}}
\vspace{-0.4cm}
\caption{CCD-based (a) precision and (b) recall of the Mask R-CNN model (blue lines) and the Ray-Tracing algorithm (orange lines). We consider both the combination of all types of artifacts (solid lines) and the combination of `Rays'+`Bright' (dashed lines).}
\label{fig: Metrics_Comparison}
\end{figure*}

\begin{figure}[!ht]
\centering
\includegraphics[width=1.0\columnwidth]{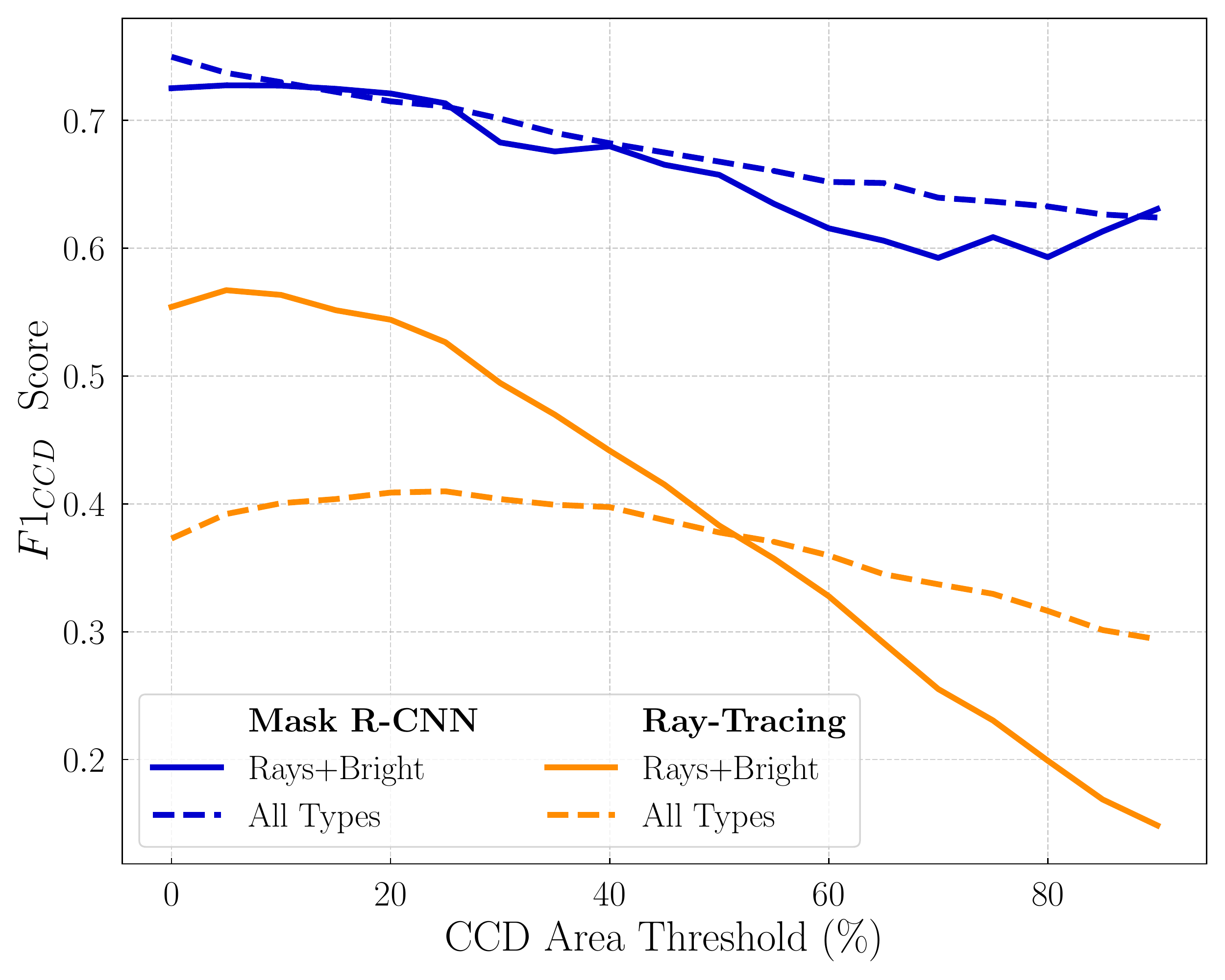}
\caption{CCD-based $F1$ scores for the same models and ghost type combinations as in Fig.~\ref{fig: Metrics_Comparison}.}
\label{Fig: F1_Comparison}
\end{figure}

In panel (a) of Fig.~\ref{Fig: Individual_metrics}, we present precision and recall as a function of the CCD area threshold for the three artifact categories individually. These metrics are related to the number of CCDs (as opposed to the number of artifacts) that were correctly or incorrectly classified. Therefore, the differences we observe between the artifact types depend on the different sizes of the artifacts. For example, as we have seen (Fig.~\ref{Fig: Ghost_Areas}), `Faint' ghosts tend to cover $\sim 10-30$ CCDs, while `Bright' ghosts are significantly smaller, covering $\sim 1-3$ CCDs. Thus, the classification and masking of a single large `Faint' object has a greater effect on the metrics than the detection of two or three `Bright' ghosts.

There are a few interesting trends to notice in this figure. First, for `Rays' and `Bright' ghosts, the precision is higher than the recall and almost constant as the area threshold changes. The high precision score ($\sim 80\%$) for these categories is easy to understand: these are the most distinct and prominent ghosts, and thus it is hard for a CCD with a `Faint' ghost (or for a CCD without a ghost) to be mistaken as containing either of these types of artifacts.

Second, the recall score for `Rays' is $\sim 70\%$ and constant as a function of the threshold. The recall score for `Bright' ghosts greatly degrades with area threshold and it is generally low (less than $50\%$). `Bright' ghosts are relatively small, only partially covering the CCDs that contain them; as we increase the area threshold, only a few such ghosts can pass it. 

A third interesting point is that `Faint' ghosts have higher recall than precision, in contrast to the two other categories. `Faint' ghosts are usually large: even though some may go undetected, the largest cover many CCDs and are usually detected (at least partially), thus pushing the CCD-based recall (completeness) to higher values. On the other hand, some `Bright' ghosts, especially those with a significant overlap with larger `Faint' ghosts can be misclassified as `Faint', leading to a lower precision. 

In panel (b) of Fig.~\ref{Fig: Individual_metrics}, we present the $F1$ score as a function of the CCD area threshold. The $F1$ score (see Eq.~\eqref{eqn:F1_score}) is  useful as a way to compare the performance of the classifier for different ghost types using a single metric. As we can see in this figure, the Mask R-CNN performs best in finding CCDs containing `Rays', while CCDs containing `Faint' ghosts are identified with higher efficiency than CCDs containing `Bright' ghosts.

In practice, we are interested in the ability of the \textit{DeepGhostBusters} Mask R-CNN to detect combinations of ghosts and scattered-light artifacts. We present the CCD-based precision and recall as a function of the area threshold in Fig.~\ref{fig: Metrics_Comparison} (panels (a) and (b), respectively); we also present the $F1$ score in Fig.~\ref{Fig: F1_Comparison} for the two combinations, `Rays'+`Bright' (solid blue lines) and `Rays'+`Bright'+`Faint' (all ghost types, dashed blue line).

We chose this combination for two reasons: first, it allows a fairer comparison with the Ray-Tracing algorithm, which is not tuned for very low-surface-brightness ghosts (see next subsection); second, for a practical application, we may not need to reject CCDs containing very faint ghosts, because these have little influence on the surface brightness of real sources and can be effectively deblended.

\subsection{Comparison with the Ray-Tracing algorithm}
\label{sec: Comparison}

Next, we compare the performance of our Mask R-CNN model in detecting ghost-containing CCDs to that of the Ray-Tracing algorithm. We note a few details of this comparison:
\begin{itemize}
    \item The test dataset consists only of images known to contain at least one ghost or scattered-light artifact. 
    \item When plotting metrics as a function of the CCD area threshold, this threshold is applied only to the ground-truth masks. This accounts for the fact that we only have predictions from the Ray-Tracing algorithm on a CCD-by-CCD basis.  
    \item The available output from the Ray-Tracing algorithm does not distinguish between the different artifact categories. Furthermore, the Ray-Tracing algorithm applies a threshold to  the predicted surface-brightness of artifacts, and thus is not optimized to detect `Faint' ghosts. For that reason we exclude `Faint' ghosts when evaluating metrics to compare performance between the Ray-Tracing and Mask R-CNN algorithms.
\end{itemize}

We plot the CCD-based precision and recall (Fig.~\ref{fig: Metrics_Comparison}) and  $F1$ score (Fig.~\ref{Fig: F1_Comparison}) resulting from the Ray-Tracing algorithm (orange lines) and Mask R-CNN (blue lines), as a function of the ground truth threshold area. We consider two categories of artifacts selected based on the ground truth masks: all ghost types combined (solid lines) and the combination of `Rays'+`Bright' ghosts (dashed line).

We first consider the limit of zero percent CCD area threshold: a single pixel of an artifact has to be in the CCD to be classified as ghost-containing. The Ray-Tracing algorithm achieves a high precision score, which, for the case when the combination of all ghost types is considered, is higher than that from the Mask R-CNN for the same case ($\sim 0.9$ vs.\ $\sim 0.7$). However, for the same case the recall is much lower ($\sim 0.8$ vs.\ $\sim 0.3$). In other words, Ray-Tracing produces results high in purity but low in completeness. When the combination of only `Rays'+`Bright' ghosts is considered, both the precision and the recall from the \textit{DeepGhostBusters} Mask R-CNN model are significantly higher than those from the Ray-Tracing algorithm.

Fig.~\ref{fig: Metrics_Comparison} shows that precision decreases, while recall increases as a function of the CCD area threshold for both artifact combinations. As we increase the threshold, fewer CCDs are labeled as containing artifacts and thus the purity decreases while the completeness increases.

The $F1$ score, which combines precision and recall, demonstrates that the performance of the Mask R-CNN model is significantly higher than that of the Ray-Tracing algorithm for all area threshold values and for both artifact combinations (Fig.~\ref{Fig: F1_Comparison}). 

\begin{table*}[!ht]
\caption{CCD-based evaluation metrics (precision, recall, $F1$ score) for the Mask R-CNN and Ray-Tracing algorithms, at $0\%$ CCD area threshold.}  
\label{table:Metrics_Comp_Table}
\centering
\begin{tabular}{|c||c|c|c|c|}
\hline
\diaghead{\theadfont Diag ColumnmnHead I}%
{\textbf{{\normalsize{Metric}}}}{{\normalsize{\textbf{Model}}}}& \multicolumn{2}{|c|}{Mask R-CNN} &  \multicolumn{2}{|c|}{Ray-Tracing}\\
\cline{2-5}
{} & Rays+Bright & Rays+Bright+Faint & Rays+Bright & Rays+Bright+Faint \\
\hline \hline
Precision & 84.3$\%$ & 68.7$\%$ & 64.7$\%$ & 89.9$\%$ \\
Recall  & 63.6$\%$ & 82.5$\%$& 48.4$\%$ & 23.5$\%$ \\
$F1$ score & 72.5 $\%$ & 75.0$\%$ & 55.4$\%$ & 37.3$\%$\\
\hline
\end{tabular}
\end{table*}

To facilitate the numerical comparison of the performance of the algorithms, we present in Table~\ref{table:Metrics_Comp_Table} the values of the different metrics for the two models, at a one pixel ($>0\%$) CCD area threshold, for both algorithms. The results for both artifact category combinations (`Rays'+`Bright' and `Rays'+`Bright'+`Faint') are presented.

\subsection{Standard object detection evaluation metrics}
\label{sec: standard_metrics}

\begin{figure*}[!ht]
\centering
\subfigure[]{\includegraphics[width=0.4\textwidth]{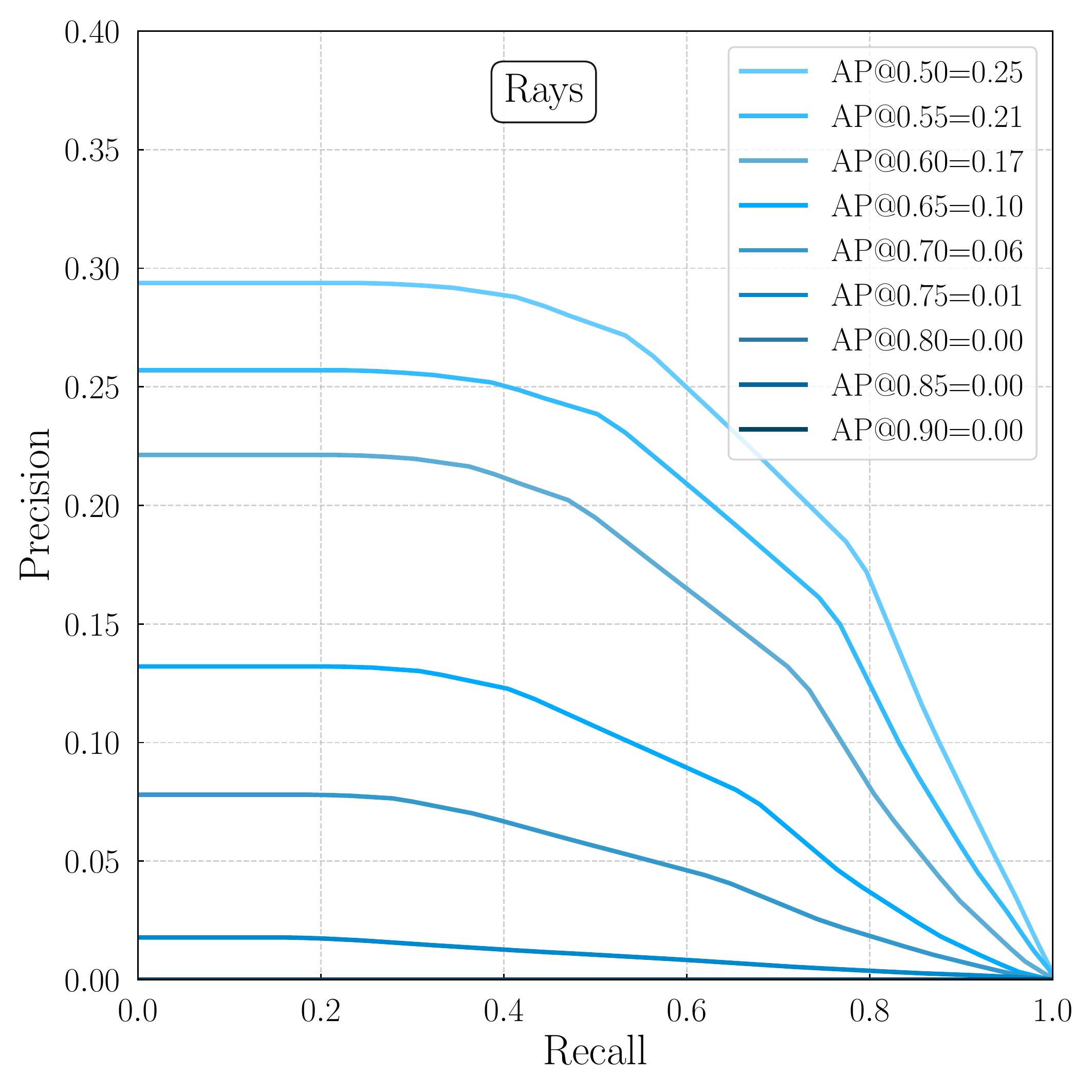}}
\hspace*{0.25cm}
\subfigure[]{\includegraphics[width=0.4\textwidth]{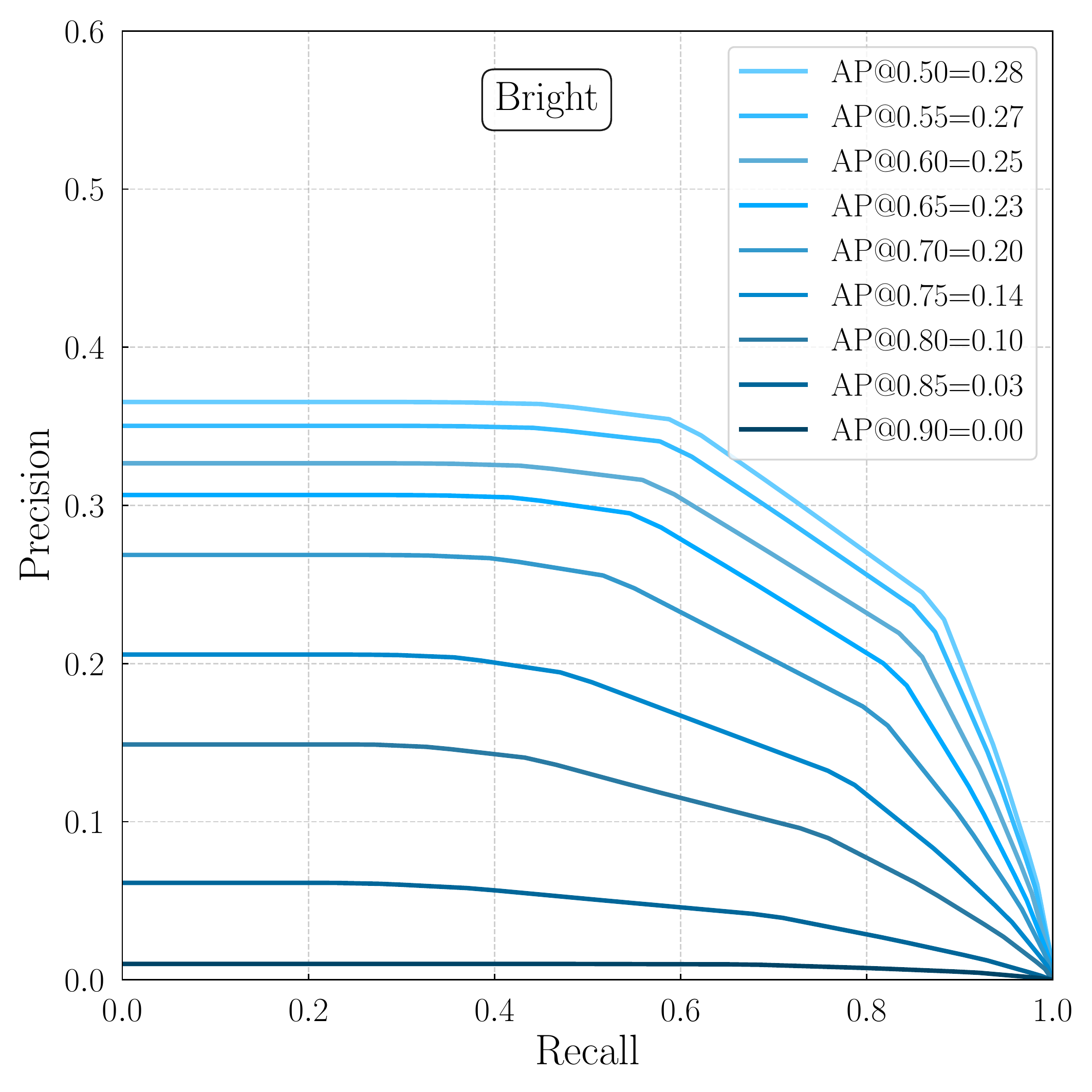}}
\vspace{0.20cm}
\subfigure[]{\includegraphics[width=0.4\textwidth]{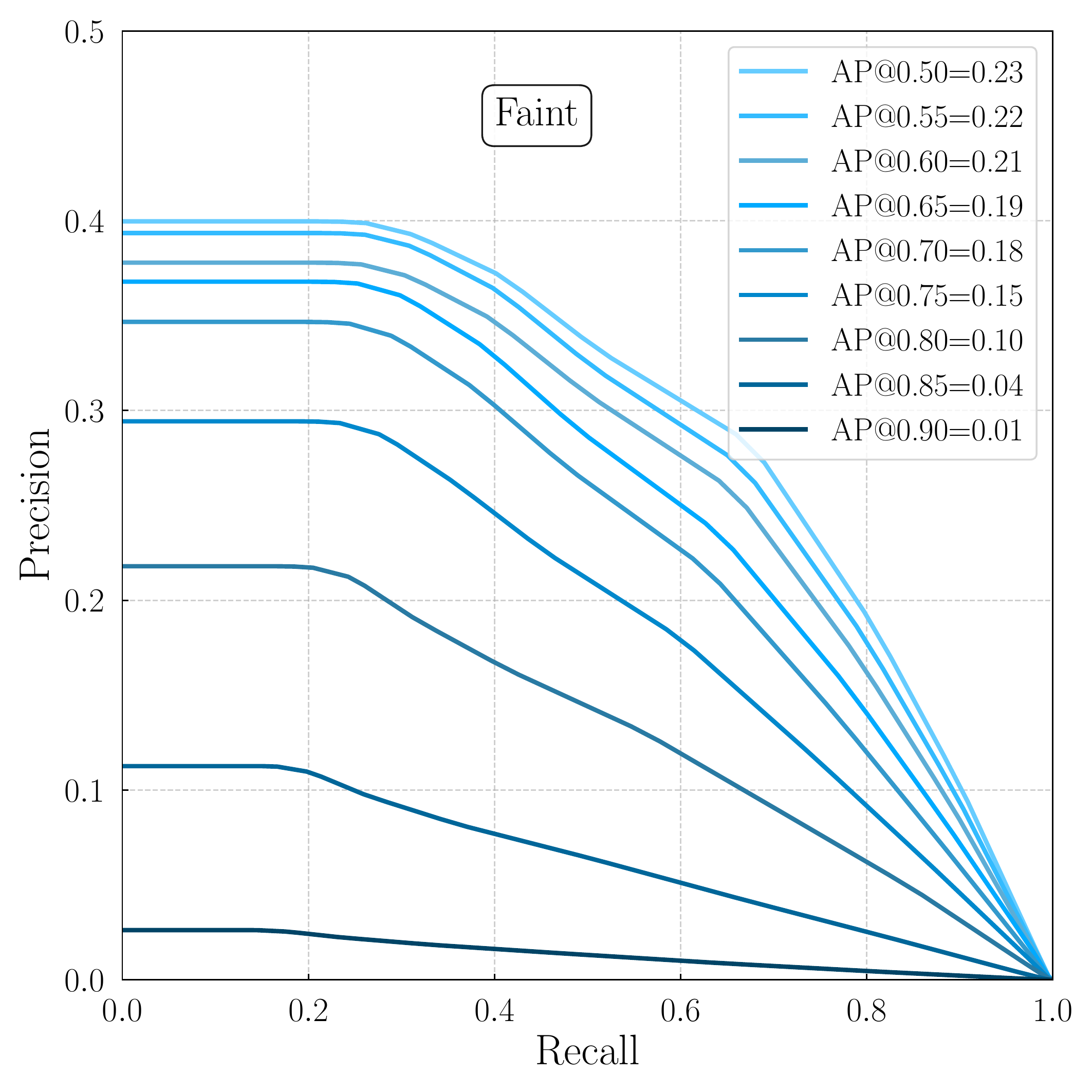}}
\hspace*{0.25cm}
\subfigure[]{\includegraphics[width=0.4\textwidth]{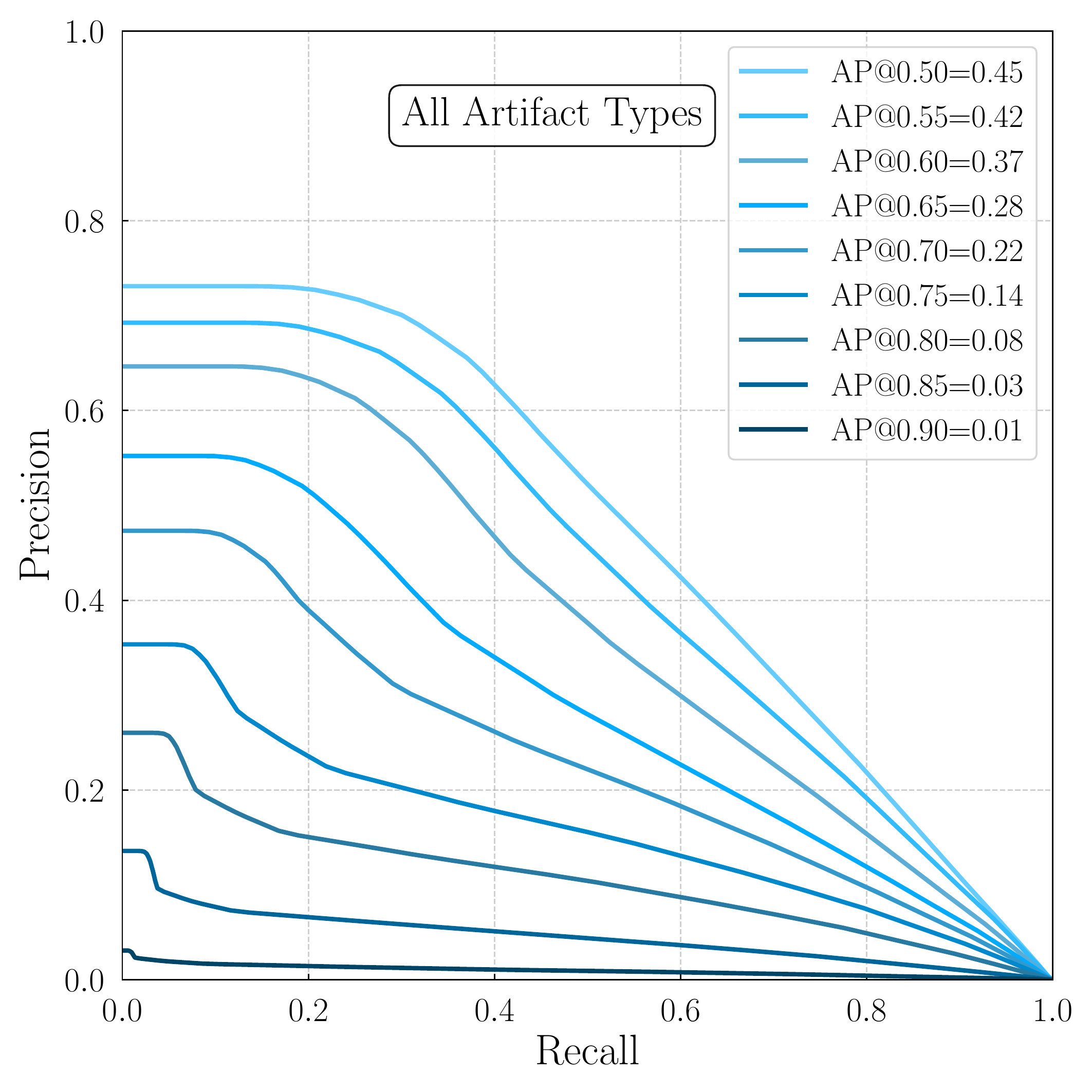}}
\vspace{-0.4cm}
\caption{Precision-Recall curves and Average Precision scores at different IoU threshold values in the range $0.50-0.90$. We show these metrics for the different ghost types in this work (`Rays'-`Bright'-`Faint'), and for all ghost types, combined. \vspace{3em}}
\label{fig: Prec_Rec_curve}
\end{figure*}

We now examine the Average Precision \citep[AP;][]{VOC_challenge}, a metric that is commonly used by the computer vision community to assess the performance of object detection algorithms. The AP is defined as the area under the Precision-Recall (PR) curve:
\begin{equation}
\mbox{AP} = \int_0^1 p(r) dr,
\end{equation}
where $p(r)$ is the precision, $p$, at recall level $r$.
In practice, an 11-point interpolation method is used, and the AP score is calculated as:
\begin{equation}
    \mbox{AP} = \frac{1}{11} \sum_{r_i\in R} \Tilde{p}(r_i),
\end{equation}
where $\Tilde{p}$ is the maximum precision at each recall bin and $R = \{0.0,0.1,\dots,1.0\}$. Precision and recall are defined using the common formulae (Eqs.~\ref{eqn:precision} and \ref{eqn:recall}), but here the number of true positives, true negatives etc. refer to detections of individual artifacts and not single CCDs.

To define the detection of an artifact, we introduce the concept of the Intersection over Union (IoU; also known as the Jaccard index; \citealt{Jaccard:1912}), which quantifies the overlap between the  masks of the ground truth and the prediction. As the name suggests, it is defined as the ratio of the area of the intersection of the predicted mask ($pm$) and the ground truth ($gt$) mask over the area of the union of the predicted and ground truth masks:
\begin{equation}
\mbox{IoU} = \frac{\mbox{area of intersection}}{\mbox{area of union}} =  \frac{area(gt \cap pm)}{area(gt \cup pm)}.
\end{equation}
An IoU threshold is then used to determine if a predicted mask is a $TP, FP$, or $FN$. It is common to evaluate the AP score at different IoU levels, and we denote the AP at a IoU threshold $\beta$ as ``AP$@ \beta$''. 

By calculating the PR curves and the AP score at different IoU threshold and for the different artifact categories, we evaluate the performance of the Mask R-CNN model for different artifact categories. Furthermore, by determining how AP varies with increasing IoU, we evaluate the agreement between the true and predicted masks. 

In Fig.~\ref{fig: Prec_Rec_curve}, we present the PR curves and the corresponding AP scores for IoU thresholds in the range $0.5-0.9$ (with step size $0.05$) for the three artifact types in panels (a)-(c), individually, and for all artifact types combined in panel (d).
We find that `Bright' ghosts are most easily detected by the Mask R-CNN, while `Faint' ghosts are the most challenging to detect --- in agreement with our expectations. Furthermore, for `Rays', the AP decreases rapidly with increasing IoU threshold: the model struggles to accurately reproduce the ground truth masks for these artifacts. This is expected, because these artifacts do not have clear boundaries, as demonstrated by variation in the mask regions defined by the human annotators. 

In that section, we have shown that the Mask R-CNN algorithm is superior to the Ray-Tracing in detecting CCDs affected by ghosts or scattered-light artifacts. 

\subsection{Using Mask R-CNN to classify ghost-containing vs.\ ghost-free images}
\label{sec: Classifier}

\begin{figure*}[!ht]
\centering
\subfigure[]{\includegraphics[width=0.9\columnwidth]{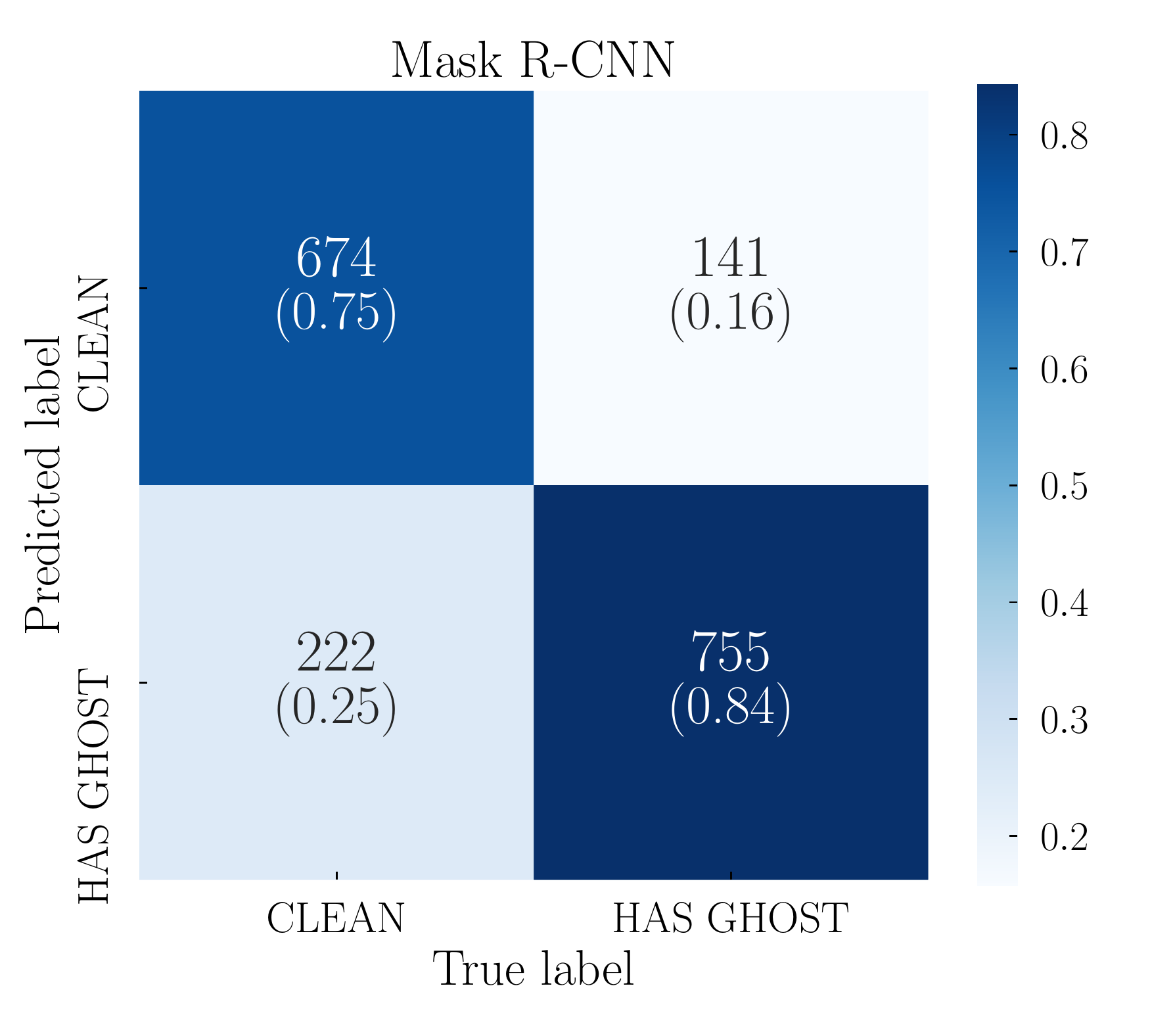}}
\hspace*{0.2cm}
\subfigure[]{\includegraphics[width=0.9\columnwidth]{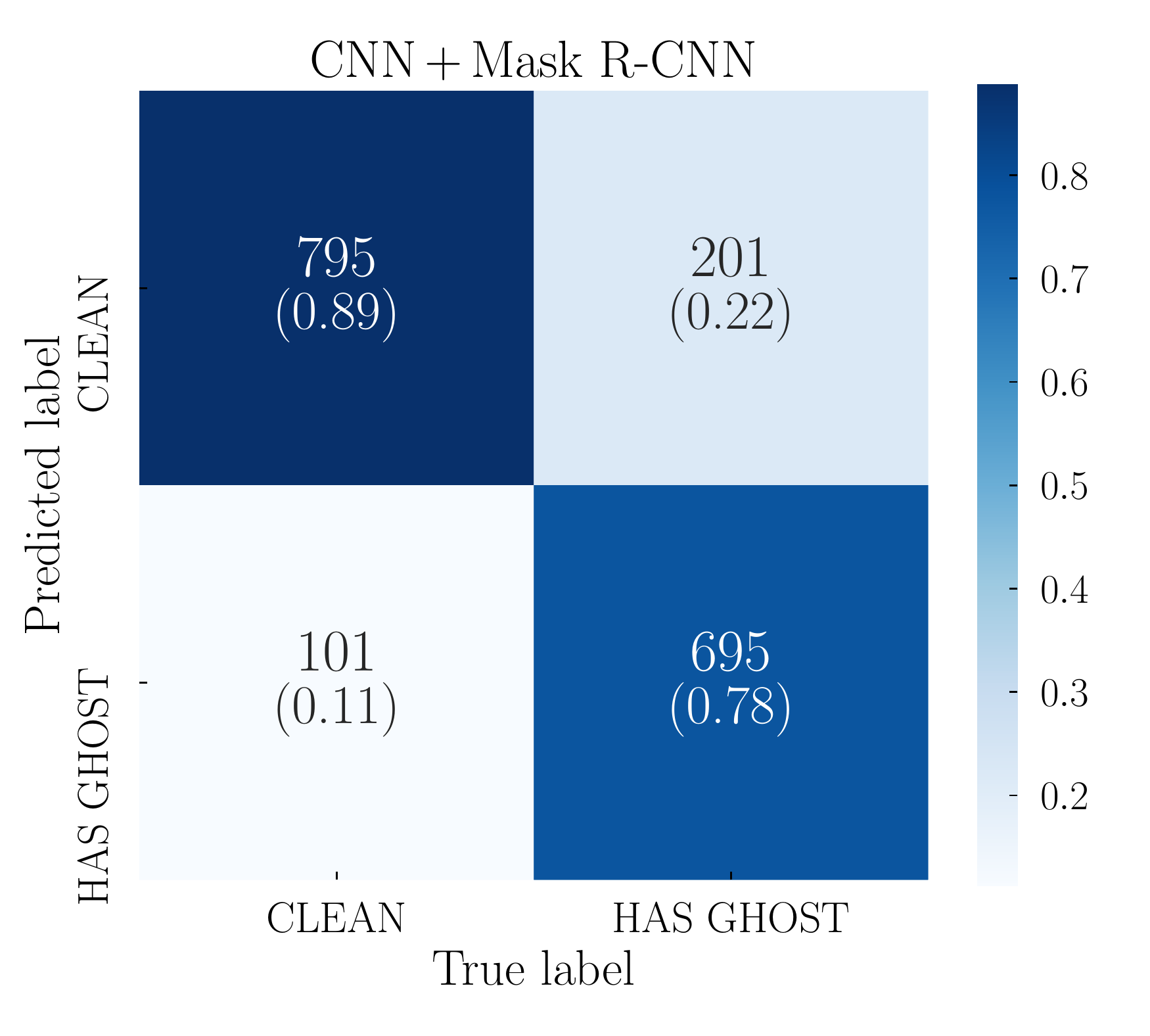}}
\vspace{-0.2cm}
\caption{(a) Confusion matrix of predictions of the Mask R-CNN model on a dataset containing an even number of ghost containing and clean images. An image is predicted to `have ghost' if even a single ghost is detected in that image by the Mask R-CNN model. (b) Confusion matrix of the predictions of the combined CNN + Mask R-CNN model (CNN model from \citet{CChang_2021}). An image is said to `have ghost' if and only if both the CNN and the Mask R-CNN models agree on that (otherwise the prediction is   `clean').}
\label{Fig: Confusion_matrix}
\end{figure*}

So far, the images used for training and testing the performance of the Mask R-CNN model were known (by visual inspection) to contain at least one ghost or scattered-light artifact. However, most DECam images do not contain prominent ghost or scattered-light artifacts, and thus they systematically differ from those used to train and test the model. 
Such differences may result in a large number of false positive detections (e.g., real astronomical sources, especially large and bright objects) or systematically failing to detect ghosts in some images --- for example, images that contain only very small or very faint ghosts. 

To test the performance of the Mask R-CNN on images that do not contain ghosts, we use a set of 1792 images with an equal number of ghost-free and ghost-containing images. This set of images is independent of the 2000 images used to train, validate, and test the Mask R-CNN model. They constitute the test set used in \citet{CChang_2021}. For this dataset, the ground truth labels refer to the presence of a ghost in the image --- not the number of ghosts or the regions affected by ghosts.

We run Mask R-CNN on this dataset: when the algorithm predicts the existence of even a single ghost or scattered-light artifact in the image, we assign a predicted label `HAS GHOST' to that image. Otherwise the assigned predicted label `CLEAN'. 
The confusion matrix resulting from this process is shown in Fig.~\ref{Fig: Confusion_matrix}. The accuracy is $79.7\%$, the precision is $77.3\%$, and the recall is $84.3\%$. Both the numbers of false positive and false negative cases are high: false positives occur at $\sim 22.7\%$ of the total number of images classified as positives, and the false negatives occur at (Dimitrios)....

However, visual inspection of false positive examples and the predicted masks revealed that most contain objects or exhibit features similar to those found in ghost-containing images. These include bright streaks from artificial Earth-orbiting satellites (mimicking `Rays'), low-surface-brightness emission from Galactic cirrus, images with poor data quality (due to cloud coverage that diffuses starlight), or large resolved stellar systems (e.g., dwarf galaxies and globular clusters). These are very similar to the cases of false positives returned by the CNN classifier in \citet{CChang_2021}.
Similarly, most of the false negatives contain very small and faint ghosts (and usually each image contains only one such ghost) that could have been easily missed even by a human annotator.\footnote{Examples of  false positives and false negatives can be found in \ref{sec: FP_FN_examples}.}
Thus, we conclude that the false positives/negatives are qualitatively different from the true positives/negatives. and that -- in practice -- the Mask R-CNN is much better in classifying images that contain unusual and/or problematic areas, compared to what one would naively assume from the confusion matrix (Fig.~\ref{Fig: Confusion_matrix}). 

We note that in practical applications of Mask R-CNN, we can reduce the number of false positives by first applying the CNN classifier presented in \citet{CChang_2021}, and then applying the Mask R-CNN only to those images that are identified as containing ghosts or scattered-light artifacts.
The results of this process on the test dataset are presented in panel (b) of Fig.~\ref{Fig: Confusion_matrix}. We find that we are able to reduce the number of false positives to less that of the Mask R-CNN alone, but at the expense of increasing the number of false negatives. This combined model has an overall accuracy of $83.1\%$, precision of $87.3\%$, and recall of $75.6\%$. Because of this trade-off, the final decision of pre-processing with a CNN depends on the particular problem and whether we are willing to reject otherwise real astronomical objects (false positives) or to have residual ghost and scattered-light artifacts (false negatives).

\section{Summary and Conclusions}
\label{sec: Summary_and_Conclusions}

In this work, we applied a state-of-the art object detection and segmentation algorithm, Mask R-CNN, to the problem of finding and masking ghosts and scattered-light artifacts in astronomical images from DECam.
The effective detection and mitigation of these artifacts is especially important for low-surface-brightness science, an important target of future surveys.\footnote{See, for example, \url{https://sites.google.com/view/lsstgsc/working-groups/low-surface-brightness-science}} 
Given the sheer volume of data generated by current and upcoming surveys, automated methods must be used for the identification of these artifacts.

In this paper, we compared the performance of the Mask R-CNN algorithm to two previous approaches, each of which has benefits and limitations. First, the conventional Ray-Tracing algorithm currently used by DES identifies individual CCDs affected by ghosting or scattered-light artifacts. This is a predictive model that does not use the actual imaging data to detect artifacts. Thus, its performance is limited by the accuracy of the optical model and external catalogs of bright stars, and it fails to detect a significant number of artifacts. Second, we compared to a relatively standard CNN \citep{CChang_2021}, which does not depend on modeling the optical processes that lead to the generation of artifacts or on external catalogs of bright astronomical objects. Furthermore, it separates ``ghost-containing" from ``clean" images with high accuracy. 
However, as a classifier, it does not identify the affected subregion(s) within the image: if used without further investigation, it can lead to the rejection of useful information from non--affected parts of the image. 

The Mask R-CNN approach presented in this work has the benefits of a deep learning approach --- i.e., it does not depend on physical modeling, except through that training data, themselves --- that can predict the locations of ghosts and scattered-light artifacts, which can be used to create CCD- and pixel-level masks of the affected region of an image.

We compare the ability of Mask R-CNN in masking affected CCDs in \textit{ghost-containing} images with that of the Ray-Tracing algorithm. We find that the Mask R-CNN model has superior performance, as measured by the $F1$ score, which is the harmonic mean of the precision (purity) and the recall (completeness). These results hold across different CCD area thresholds and for the two combinations of the morphological classes discussed in this work --- `Bright'+`Rays' and `Bright'+ `Rays'+`Faint'. At the threshold of one pixel ($> 0\%$), for example, and for the combination `Rays+Bright' the $F1$ score of the Mask R-CNN model is $72.5\%$ as opposed to $55.4\%$ of the Ray-Tracing algorithm.

One weakness of our method is that it produces a large number of false positives when presented with images that do not contain ghosts or scattered-light artifacts --- although many of these false positives contain other types of artifacts or bright astronomical objects. We show that, to mitigate this problem, a CNN classifier similar to that discussed in \citet{CChang_2021} can be used as a pre-processing step before the Mask R-CNN is applied to images that are predicted to contain ghosts or scattered-light artifacts. This process reduces the number of false positives by a factor of two and increases the number of false negatives, and improves the accuracy.

The results presented here highlight the promise of object detection and segmentation methods in tackling the identification of ghosts and scattered-light artifacts. Since deep learning models that are trained on one data set can be adapted to a new data set with many fewer examples through transfer learning, the \textit{DeepGhostBusters} algorithm trained on DECam images can potentially be adapted and retrained to identify such artifacts in future surveys. Indeed, cross-survey transfer learning has already been shown to significantly reduce the need for large annotated datasets in deep learning-based classification cases \citep[e.g.,][]{Domingues:2019,Khan2019,Tanoglidis_2021b}.
Additionally, these results indicate that such techniques are also promising for different, but related, problems, such as the the detection of artifacts from cosmic rays, satellite trails, etc. \citep[e.g.,][]{Goldstein:2015,Desai:2016,Melchior:2016,Zhang:2019,Roman:2020,Paillassa:2020}.
Such automated techniques can facilitate the efficient separation of artifacts from scientifically useful data in upcoming surveys like LSST. 

\section*{Acknowledgements}

We would like to thank Colin Burke, Chihway Chang, Tom Diehl, Brenna Flaugher, and Steve Kent for useful discussions and suggestions. This paper has gone through internal review by the DES collaboration.

A. \'Ciprijanovi\'c is partially supported by the High Velocity Artificial Intelligence grant as part of the Department of Energy High Energy Physics Computational HEP sessions program.

We acknowledge the Deep Skies Lab as a community of multi-domain experts and collaborators who've facilitated an environment of open discussion, idea-generation, and collaboration. This community was important for the development of this project.

This material is based upon work supported by the National Science Foundation under Grant No.\ AST-2006340.
This work was supported by the University of Chicago and the Department of Energy under section H.44 of Department of Energy Contract No.\ DE-AC02-07CH11359 awarded to Fermi Research Alliance, LLC. 
This work was partially funded by Fermilab LDRD 2018-052. 

This project used public archival data from the Dark Energy Survey (DES).
Funding for the DES Projects has been provided by the U.S. Department of Energy, the U.S. National Science Foundation, the Ministry of Science and Education of Spain, 
the Science and Technology Facilities Council of the United Kingdom, the Higher Education Funding Council for England, the National Center for Supercomputing 
Applications at the University of Illinois at Urbana-Champaign, the Kavli Institute of Cosmological Physics at the University of Chicago, 
the Center for Cosmology and Astro-Particle Physics at the Ohio State University,
the Mitchell Institute for Fundamental Physics and Astronomy at Texas A\&M University, Financiadora de Estudos e Projetos, 
Funda{\c c}{\~a}o Carlos Chagas Filho de Amparo {\`a} Pesquisa do Estado do Rio de Janeiro, Conselho Nacional de Desenvolvimento Cient{\'i}fico e Tecnol{\'o}gico and 
the Minist{\'e}rio da Ci{\^e}ncia, Tecnologia e Inova{\c c}{\~a}o, the Deutsche Forschungsgemeinschaft and the Collaborating Institutions in the Dark Energy Survey. 

The Collaborating Institutions are Argonne National Laboratory, the University of California at Santa Cruz, the University of Cambridge, Centro de Investigaciones Energ{\'e}ticas, 
Medioambientales y Tecnol{\'o}gicas-Madrid, the University of Chicago, University College London, the DES-Brazil Consortium, the University of Edinburgh, 
the Eidgen{\"o}ssische Technische Hochschule (ETH) Z{\"u}rich, 
Fermi National Accelerator Laboratory, the University of Illinois at Urbana-Champaign, the Institut de Ci{\`e}ncies de l'Espai (IEEC/CSIC), 
the Institut de F{\'i}sica d'Altes Energies, Lawrence Berkeley National Laboratory, the Ludwig-Maximilians Universit{\"a}t M{\"u}nchen and the associated Excellence Cluster Universe, 
the University of Michigan, NSF's NOIRLab, the University of Nottingham, The Ohio State University, the University of Pennsylvania, the University of Portsmouth, 
SLAC National Accelerator Laboratory, Stanford University, the University of Sussex, Texas A\&M University, and the OzDES Membership Consortium.

Based in part on observations at Cerro Tololo Inter-American Observatory at NSF's NOIRLab (NOIRLab Prop. ID 2012B-0001; PI: J. Frieman), which is managed by the Association of Universities for Research in Astronomy (AURA) under a cooperative agreement with the National Science Foundation.

The DES data management system is supported by the National Science Foundation under Grant Numbers AST-1138766 and AST-1536171.
The DES participants from Spanish institutions are partially supported by MICINN under grants ESP2017-89838, PGC2018-094773, PGC2018-102021, SEV-2016-0588, SEV-2016-0597, and MDM-2015-0509, some of which include ERDF funds from the European Union. IFAE is partially funded by the CERCA program of the Generalitat de Catalunya.
Research leading to these results has received funding from the European Research
Council under the European Union's Seventh Framework Program (FP7/2007-2013) including ERC grant agreements 240672, 291329, and 306478.
We  acknowledge support from the Brazilian Instituto Nacional de Ci\^encia
e Tecnologia (INCT) do e-Universo (CNPq grant 465376/2014-2).

This manuscript has been authored by Fermi Research Alliance, LLC under Contract No. DE-AC02-07CH11359 with the U.S. Department of Energy, Office of Science, Office of High Energy Physics.

\appendix

\section{Human annotator agreement}
\label{sec: Agreement_between_annot}

As mentioned in Sec.~\ref{sec: Annotation_pr}, human annotators do not always agree on the mask boundaries and the artifact types. A significant disagreement may affect the performance of the Mask R-CNN, so we study extent of the disagreement in more detail, which may suggest avenues for improvement of the annotation process.  

\begin{figure*}[!ht]
\centering
\subfigure[]{\includegraphics[width=0.4\textwidth]{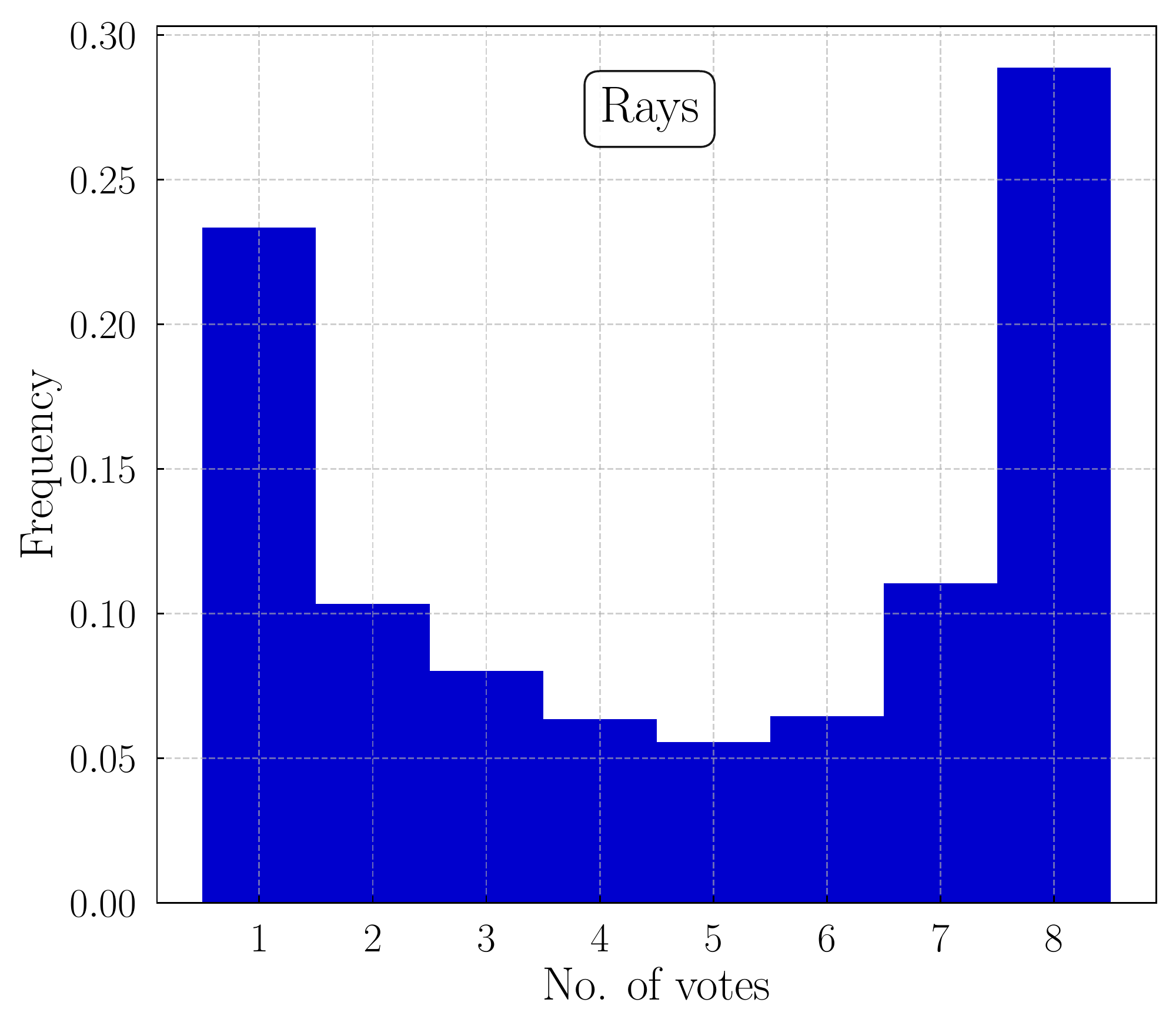}}
\hspace*{0.15cm}
\subfigure[]{\includegraphics[width=0.4\textwidth]{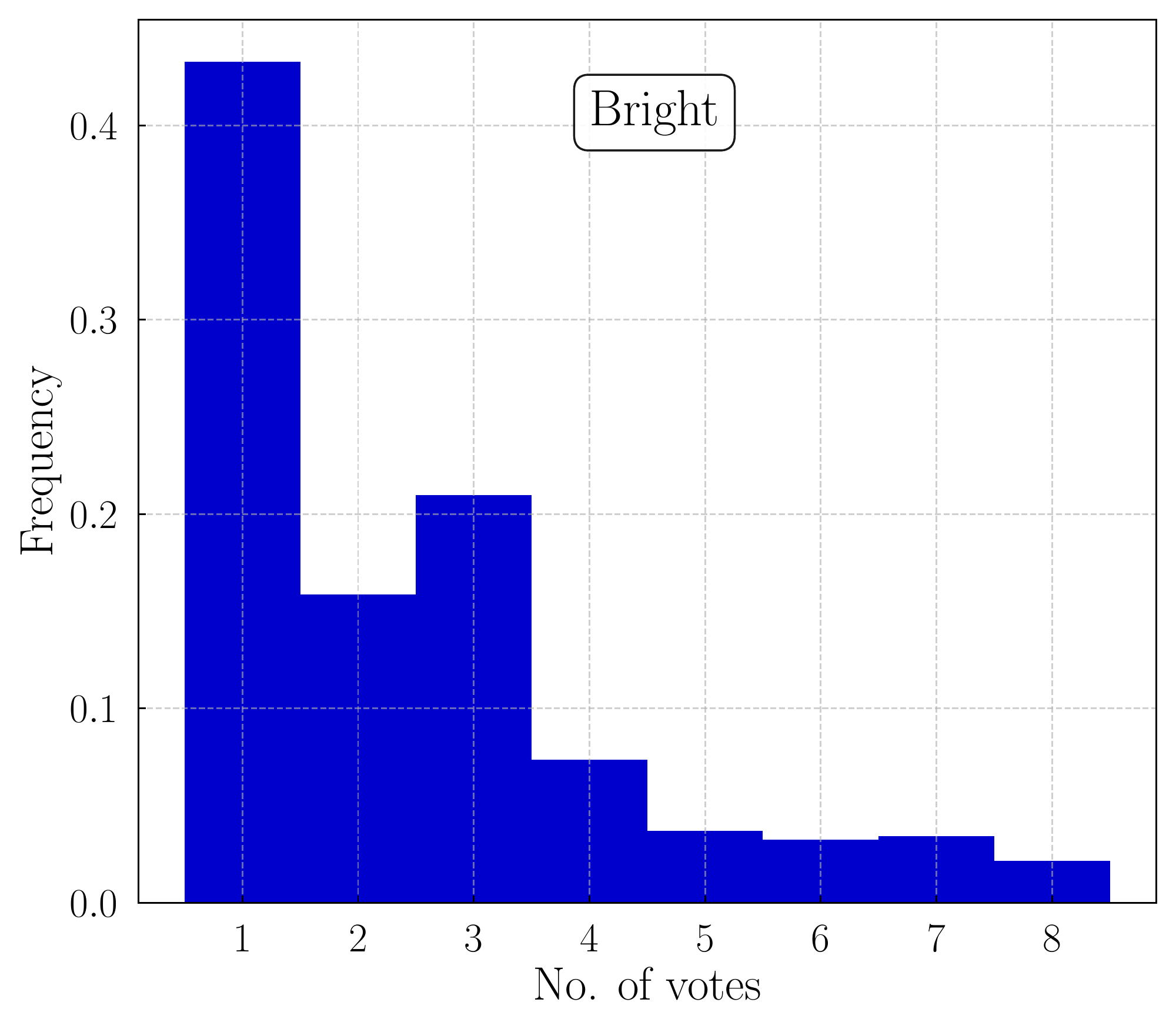}}
\subfigure[]{\includegraphics[width=0.4\textwidth]{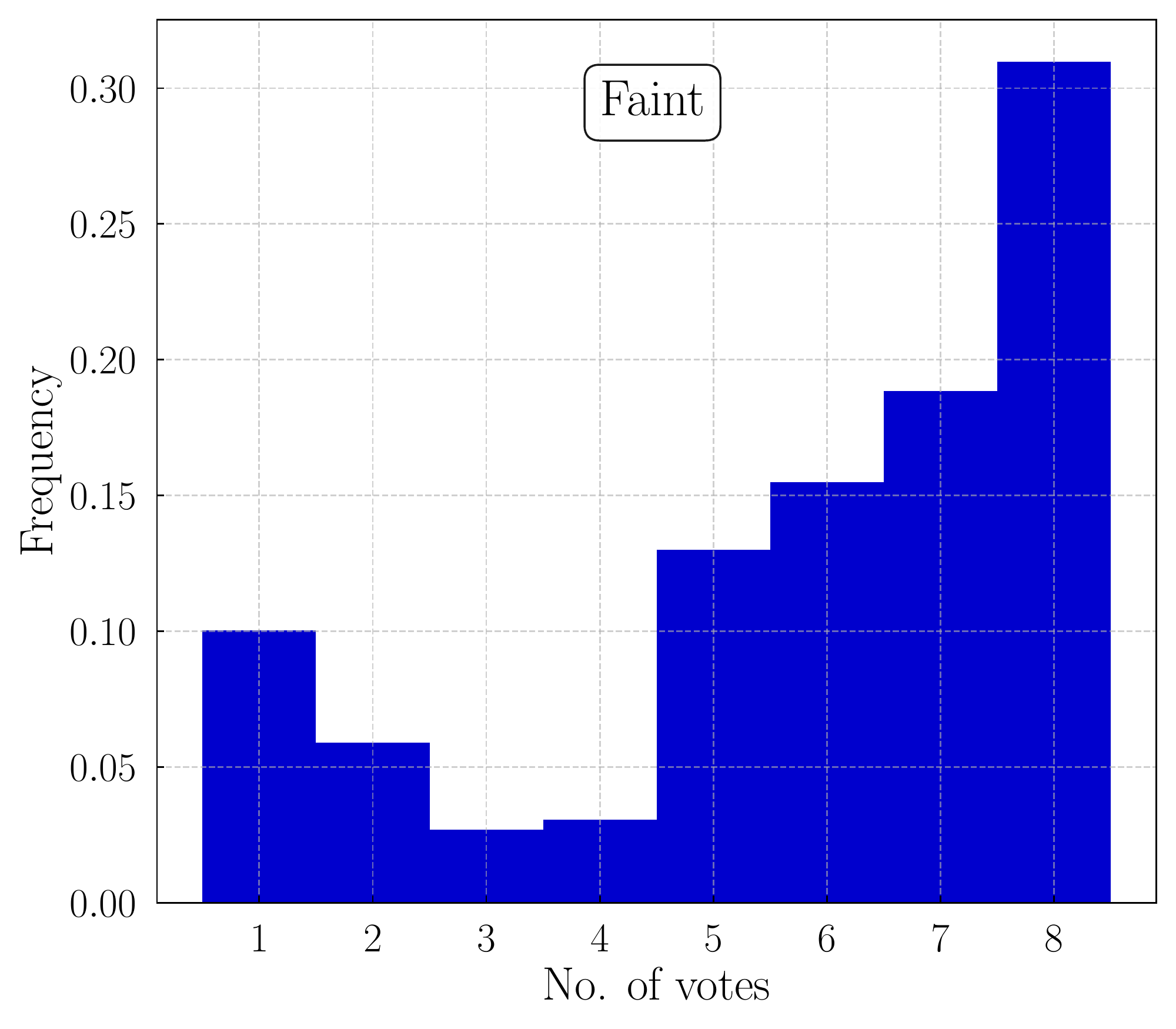}}
\hspace*{0.15cm}
\subfigure[]{\includegraphics[width=0.4\textwidth]{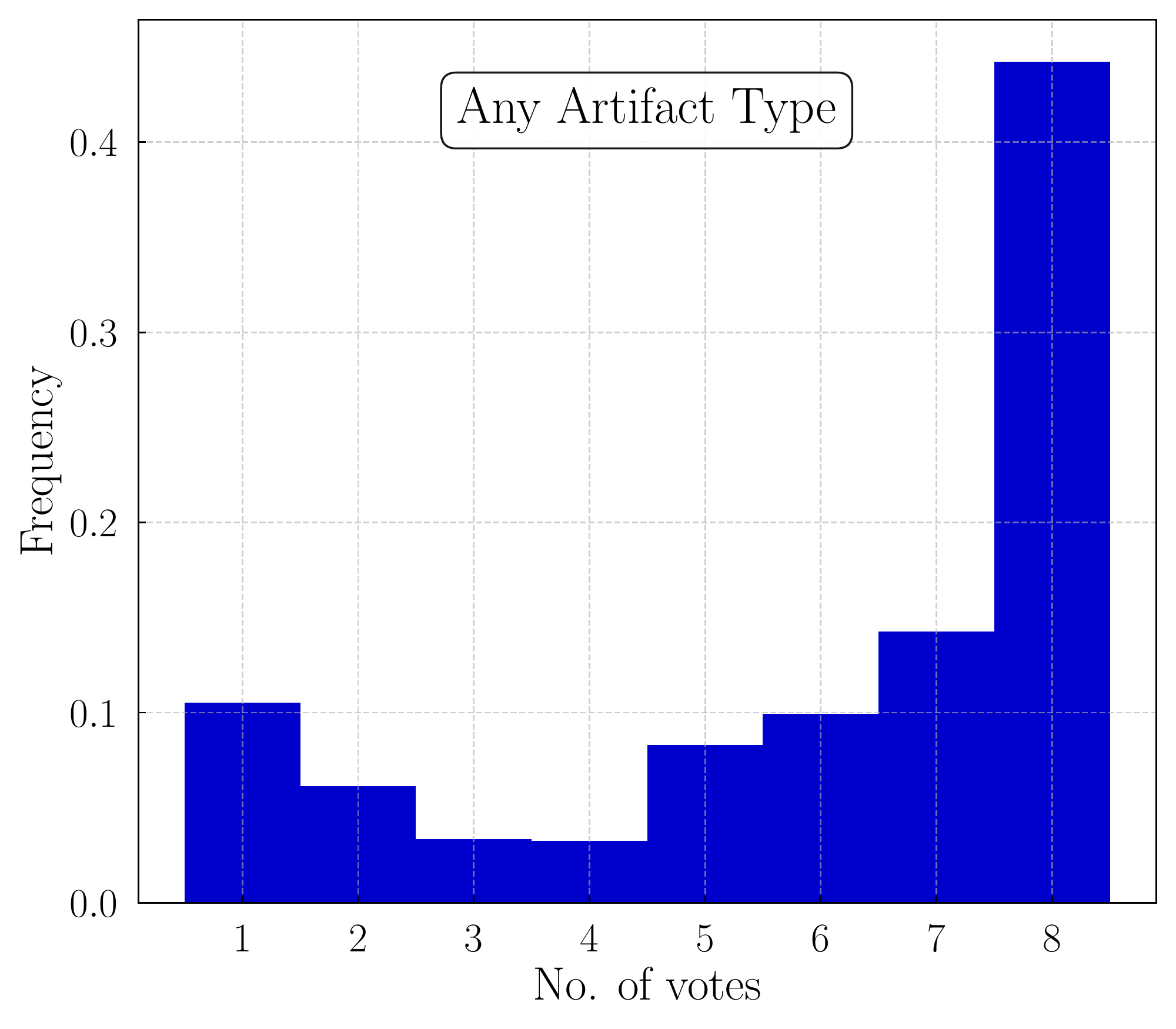}}
\vspace{-0.4cm}
\caption{Distribution of the number of votes each pixel in our dataset has received as containing a ghost, from the eight annotators. We include only pixels that have received at least one vote. We present the distributions for each ghost type separately (panels (a)-(c)) and without distinguishing between the different types (panel (d)).}
\label{fig: Agreement}
\end{figure*}

All eight annotators were given a common subset of 50 images that were randomly drawn from the full dataset described in Sec.~\ref{sec: Images}. When an annotator creates a mask for a specific artifact, they give a `vote' to the region covered by that mask. A second annotator will create a different mask around the same object. The pixels where there is an overlap between the two masks will receive two votes in total while the non-overlapping parts only one. The same process continues for all the eight annotators. The same region may receive multiple different classifications (e.g., votes for both `Bright' and `Faint' ghosts).

In Fig.~\ref{fig: Agreement}, we present histograms of the distribution of the number of votes each pixel in the dataset received during the annotation process. We restrict it to pixels that have received at least one vote. We present the distributions for each artifact category separately in panels (a)-(c), and the case where we do not distinguish between different types in panel (d). A distribution that has a strong peak in the region of $\sim 8$ votes indicates that there is a very good agreement between the annotators.

The histogram for `Rays' shows a strong bimodality, with many pixels receiving 8 votes , but also many pixels receive just 1--2 votes. These artifacts are distinct and bright, and hard to confuse with any one of the other two types. However, they do not have very clear boundaries, so, while annotators agree on the bulk of the pixels affected by a ghost, they do not agree on the extent/edges of the masks they create. 

The histogram of votes for `Bright' artifacts, panel (b), presents a peak at the low end (1--3 votes). This can be explained by the fact that there is significant confusion about the class of some large ghosts, which most annotators classify as `Faint', while a few classify as `Bright'. Since they are much larger compared to other typical `Bright' ghosts, the distribution is dominated by the pixels belonging to these confusing artifacts. 

Generally, there is a good agreement between the annotators when it comes to `Faint' ghosts, with over $30\%$ of the pixels having received the full eight votes. 
When not distinguishing between the different types of artifacts (panel (d)), we see very good agreement between the annotators in masking ghost-containing pixels, with $\sim 45\%$ of those pixels having received the maximum 8 votes, and an additional $\sim 25\%$ having received seven votes. Only $\sim 10\%$ of the pixels have received only one vote. 

From the above discussion, we conclude that there is generally good agreement in the mask-creation process. Some confusion exists between `Faint' and `Bright' ghost types, because the distinction between the two is quite arbitrary. Some potential avenues for improvement are to consider these two categories as one, define more specific criteria for each class, or have multiple persons annotate the same images and assign each artifact to the class that receives the most votes.
 
\section{Training History}
\label{sec: Losses}

\begin{figure}[!ht]
\centering
\includegraphics[width=1.0\columnwidth]{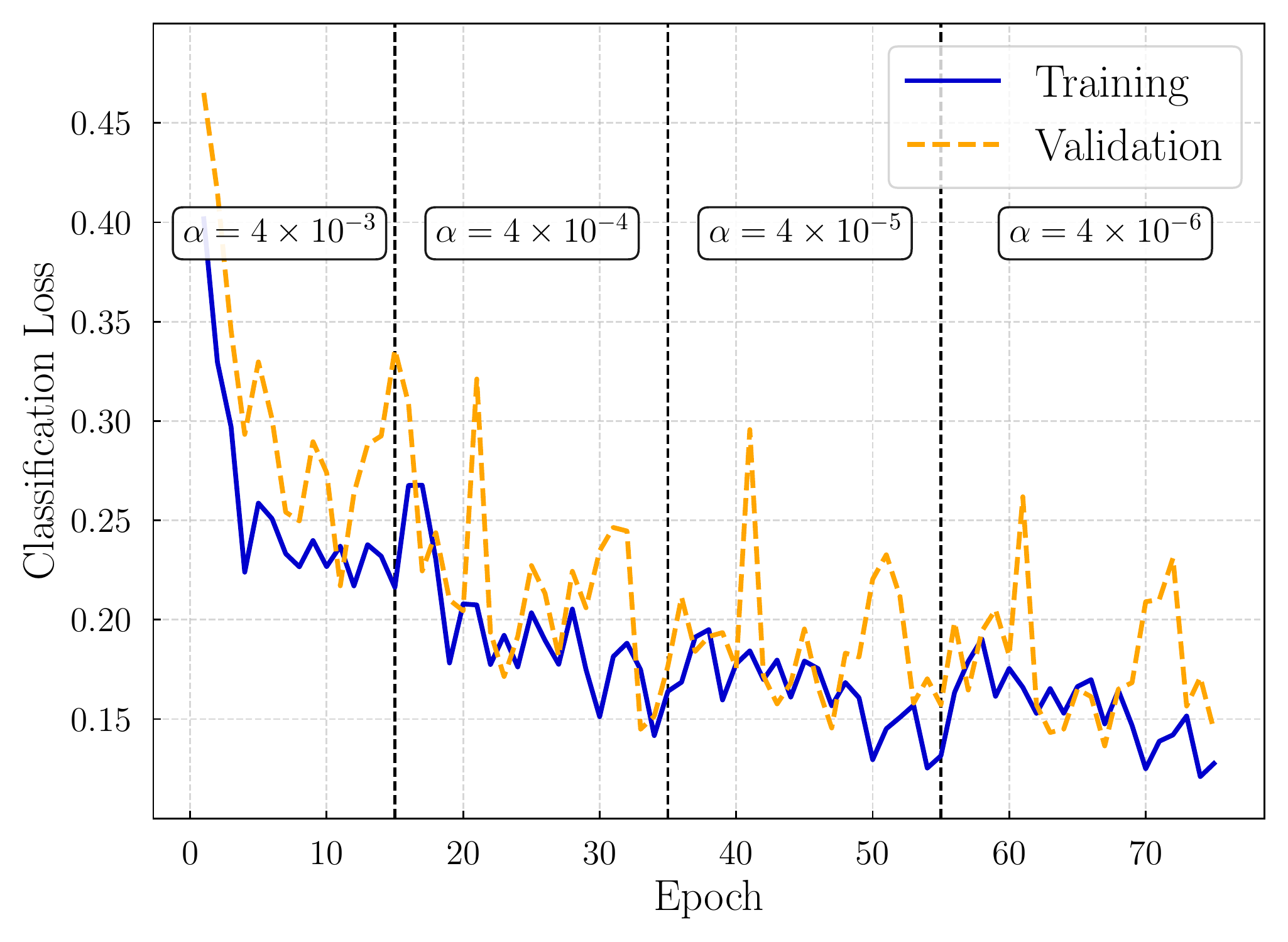}
\caption{Classification loss as a function of the training epoch.}
\label{Fig: Class_loss}
\end{figure}

\begin{figure}[!ht]
\centering
\includegraphics[width=1.0\columnwidth]{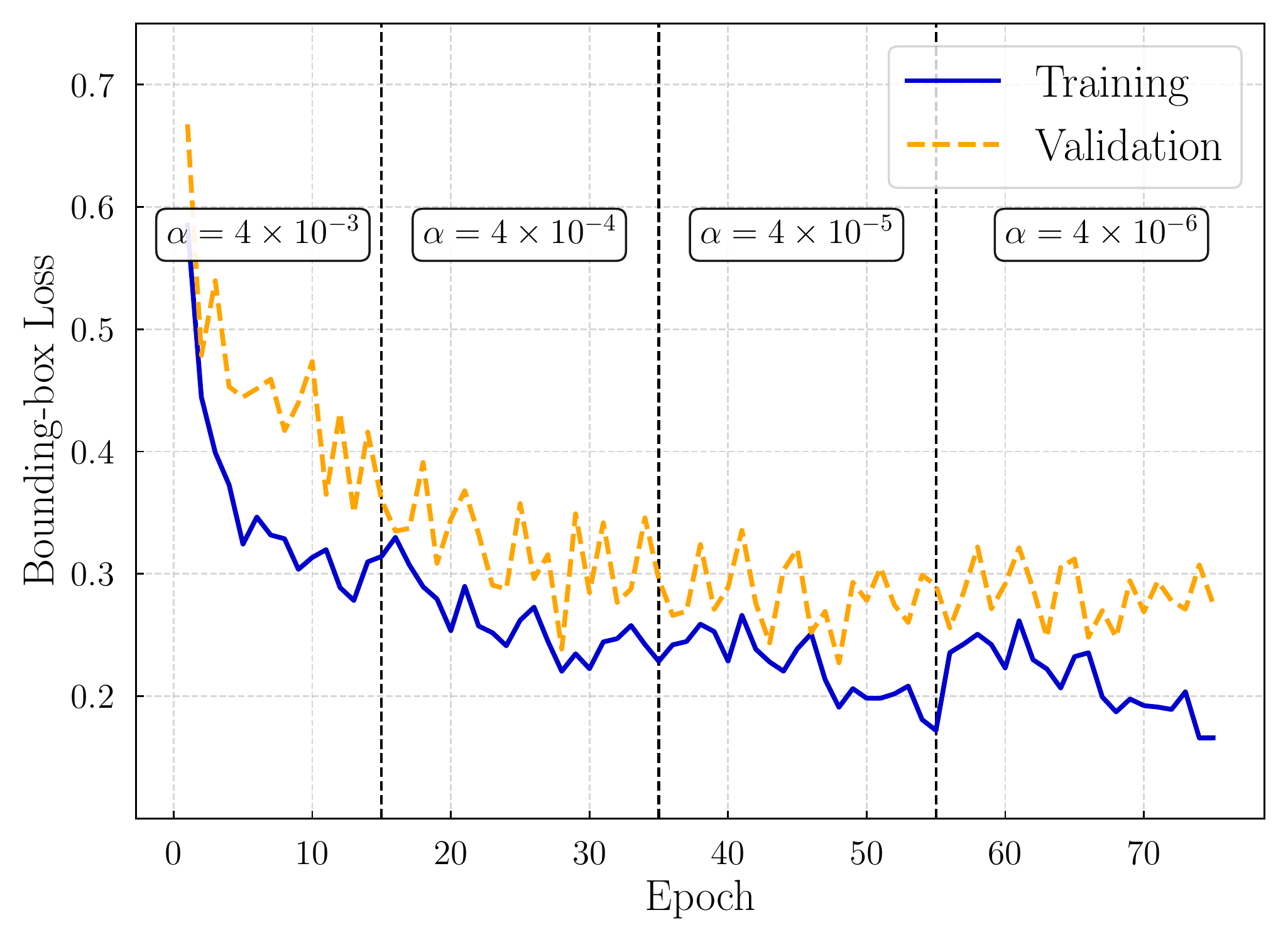}
\caption{Bounding box loss as a function of the training epoch.}
\label{Fig: bbox_loss}
\end{figure}

\begin{figure}[!ht]
\centering
\includegraphics[width=1.0\columnwidth]{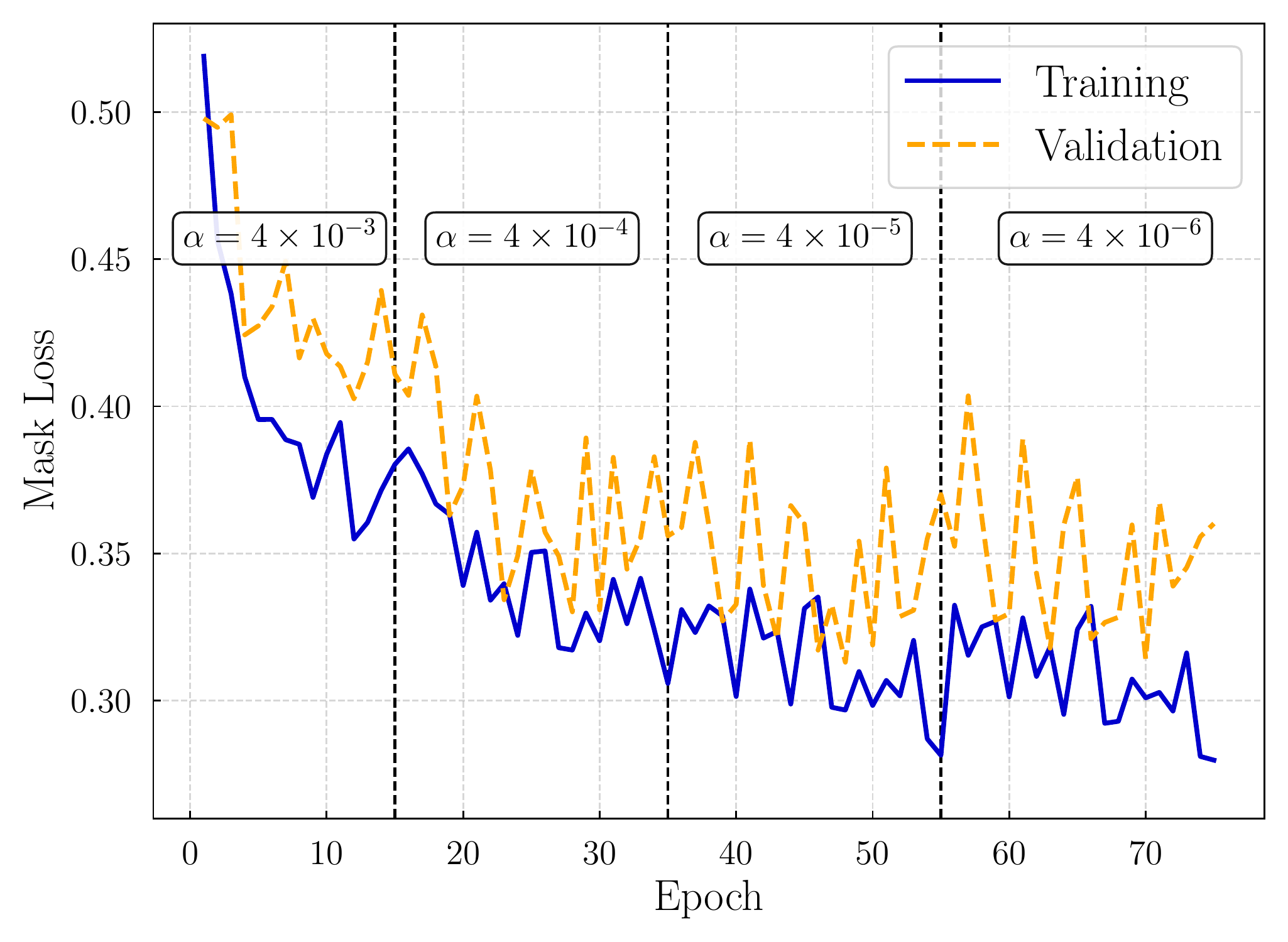}
\caption{Mask loss as a function of the training epoch.}
\label{Fig: Mask_loss}
\end{figure}

In Fig.~\ref{Fig: Total_Loss}, we presented the total loss as a function of the training epoch (training history). The total loss, $L_{\rm tot}$, is the sum of the classification, bounding box, and mask loss (see Sec.~\ref{sec: Methods}). We present the training histories for these losses individually in Figs.~\ref{Fig: Class_loss}, \ref{Fig: bbox_loss}, and \ref{Fig: Mask_loss}, respectively. As described in the main text, we train the model using progressively smaller learning rates for a finer tuning of the parameters. We stopped the process at 75 epochs due to overfitting thereafter.

\section{Masking CCDs}
\label{sec: masking}

\begin{figure*}[!ht]
\centering
\subfigure[]{\includegraphics[width=0.4\textwidth]{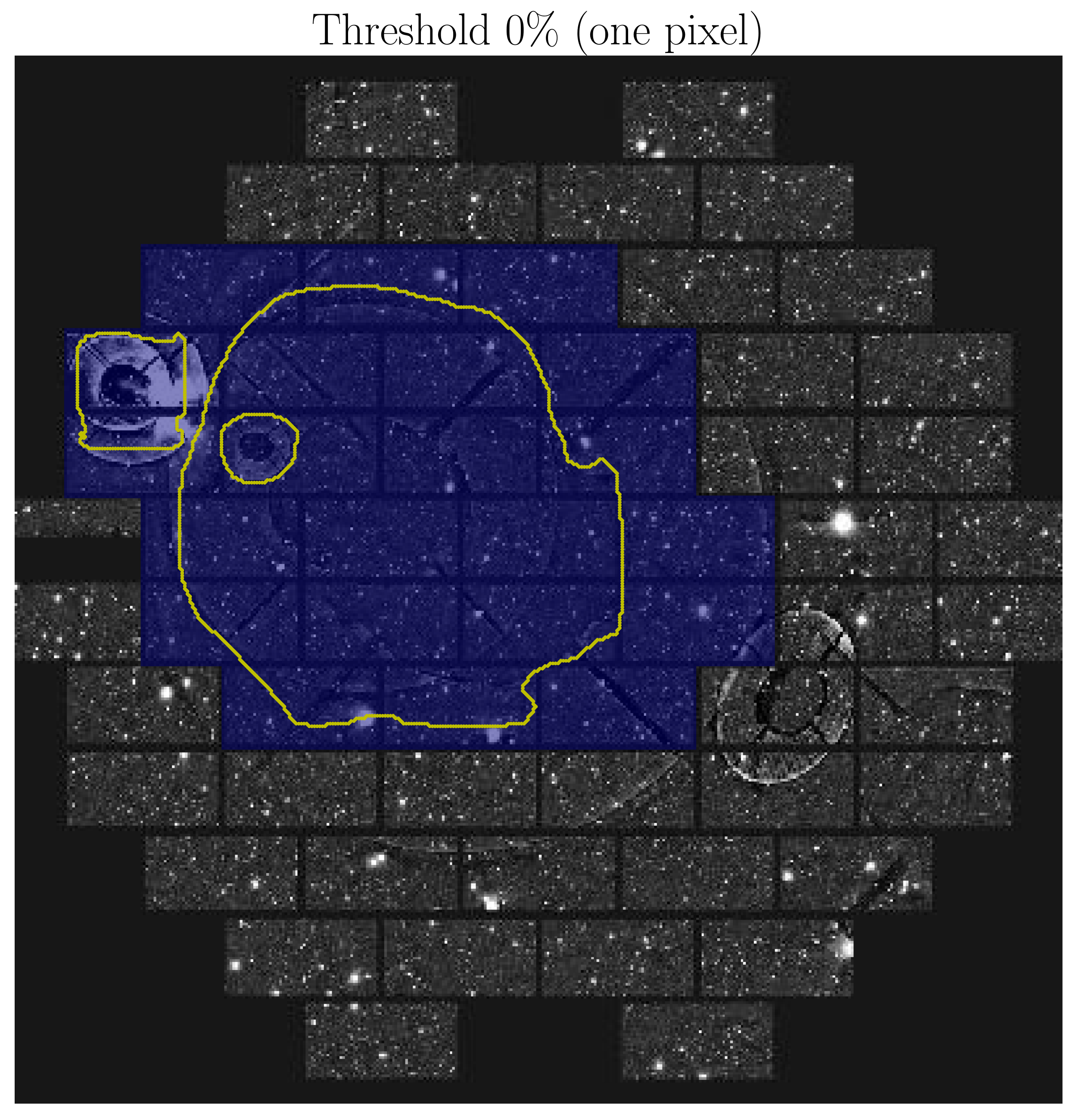}}
\hspace*{0.15cm}
\subfigure[]{\includegraphics[width=0.4\textwidth]{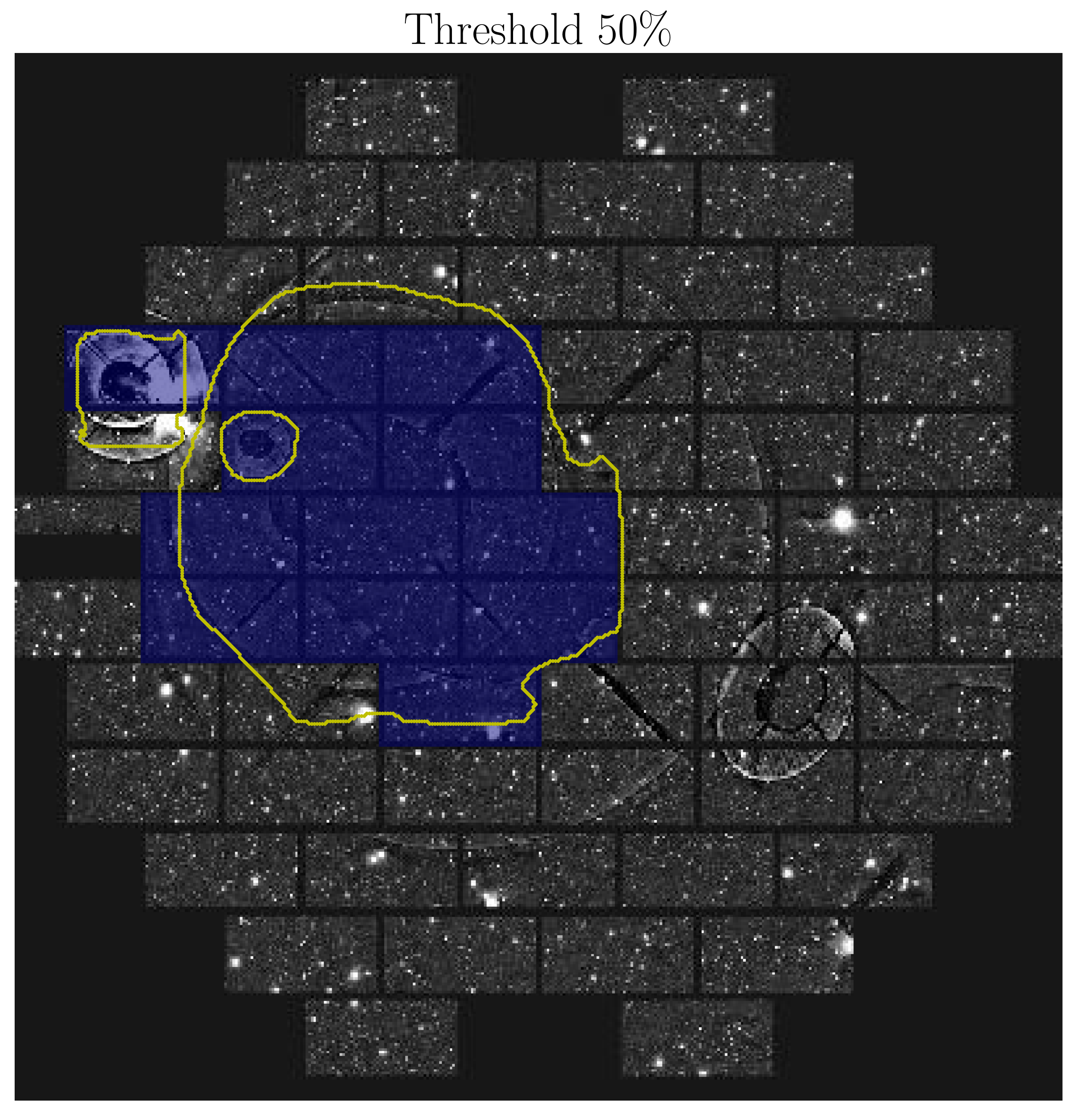}}
\subfigure[]{\includegraphics[width=0.4\textwidth]{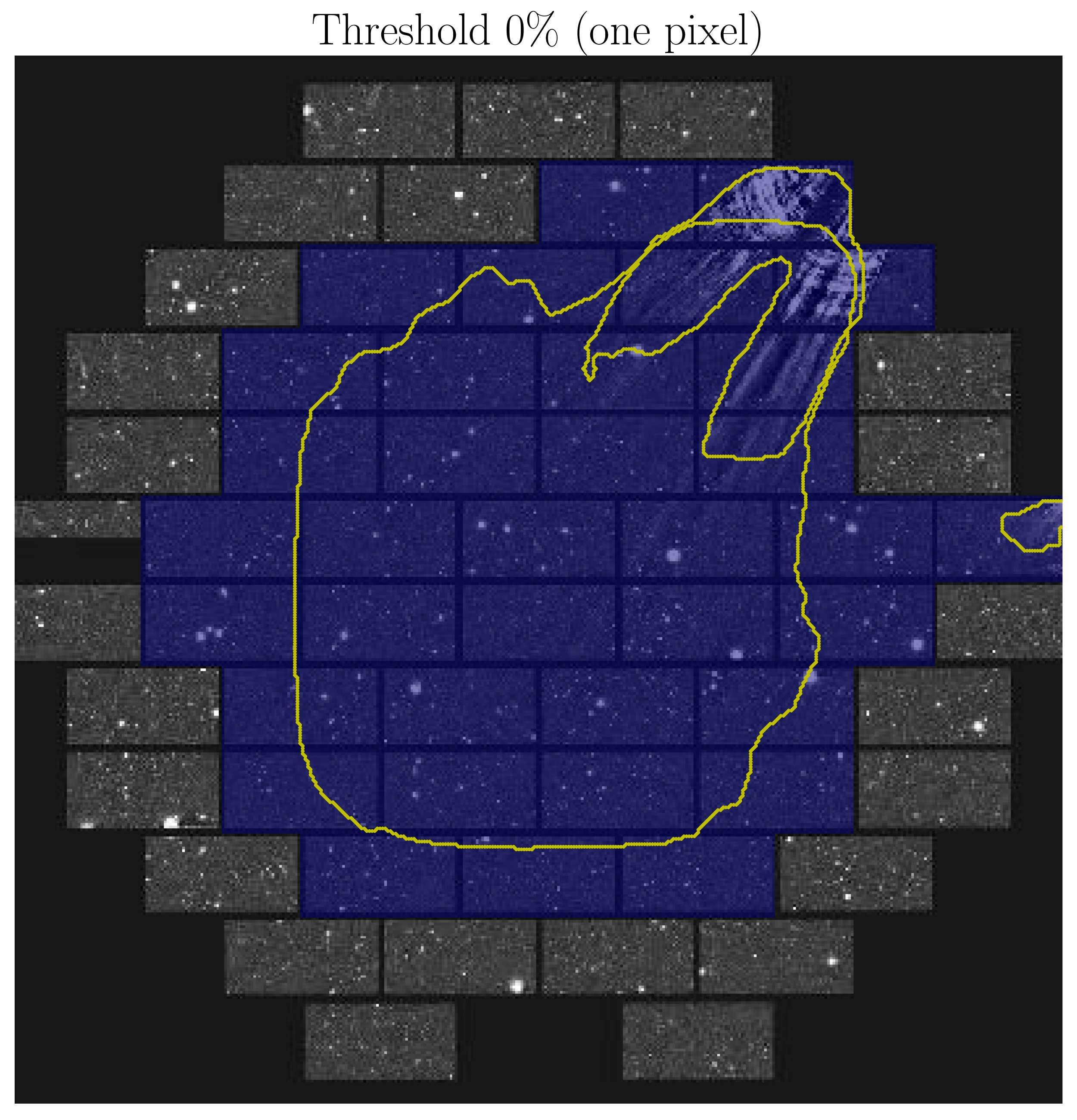}}
\hspace*{0.15cm}
\subfigure[]{\includegraphics[width=0.4\textwidth]{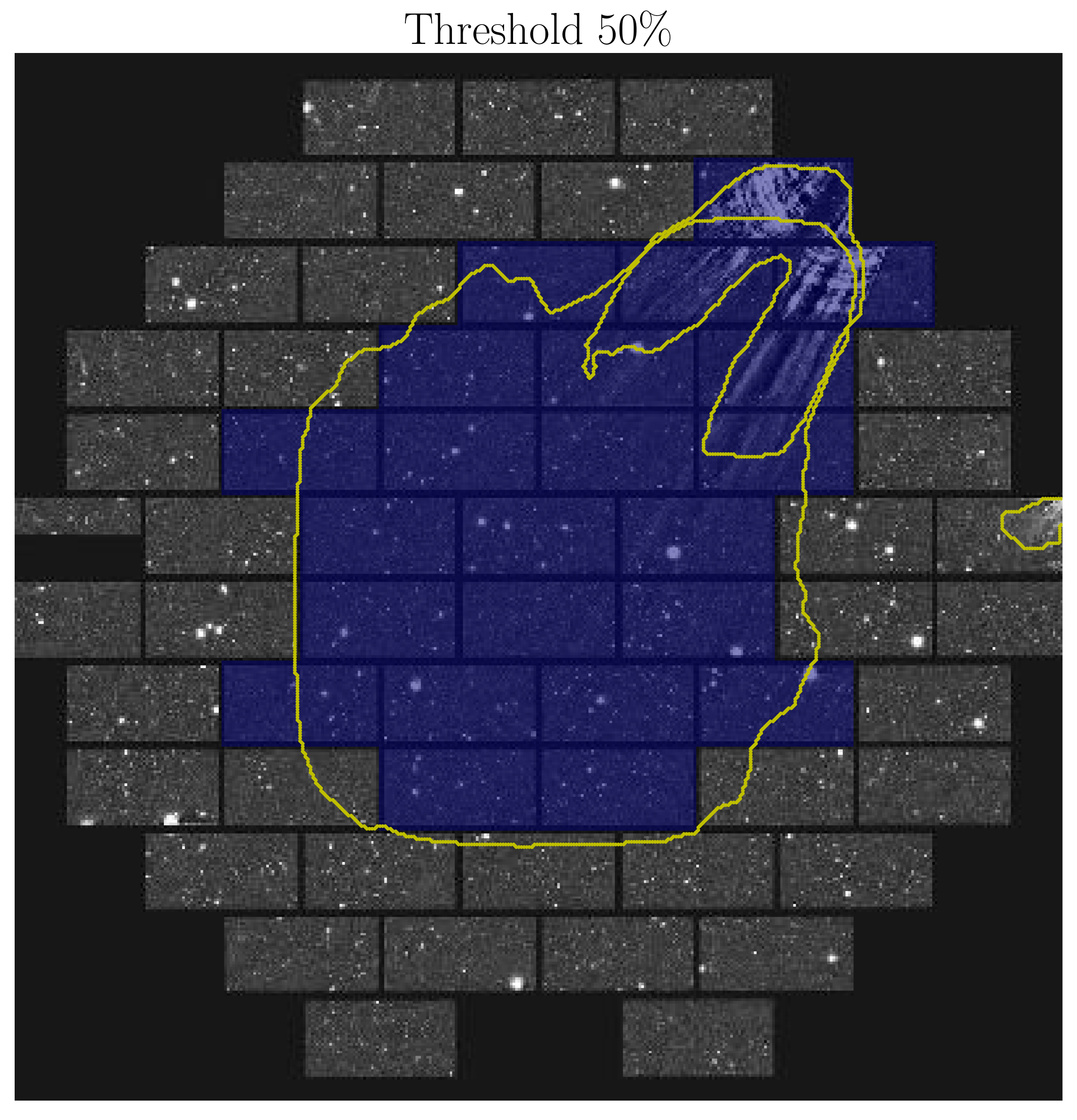}}
\vspace{-0.4cm}
\caption{CCDs masked as ghost-containing (in blue) when even a single pixel of the predicted ghost mask lies within the CCD ($0\%$ threshold, panels (a) and (c)), and when at least half of the CCD area has to be covered by the CCD (50\% threshold, panels (b) and (d)). The yellow contours correspond to the mask predictions of the Mask R-CNN model (without distinguishing between the different types of artifacts).}
\label{fig: Masks_thres}
\end{figure*}

To help the reader better understand how the imposed area threshold affects the number of CCDs classified as ghost-containing (Sec.~\ref{sec: CCD-based}), in this Appendix we present the predicted artifact masks and the affected CCDs for two different threshold levels, for the same images presented in the top row of Fig.~\ref{fig: Prediction_Examples}. 

Specifically, in the panels (a) and (c) of Fig.~\ref{fig: Masks_thres} we map (in blue) those CCDs that are classified as ghost-containing when even a single pixel of the predicted artifact mask lies within that CCD ($> 0\%$ threshold). In panels (b) and (d) we show, for the same images, the CCDs masked as ghost-containing when at least half of area of the CCD has to be covered by an artifact to be classified as such ($50\%$ threshold). To make the comparison easier, we overlay (yellow contours) the mask predictions of the Mask R-CNN model, without distinguishing between the different ghosting and scattered-light artifact types.

\section{False Positive and False Negative examples}
\label{sec: FP_FN_examples}

\begin{figure*}[!ht]
\centering
\vspace{-2.2cm}
\subfigure[]{\includegraphics[width=0.8\textwidth]{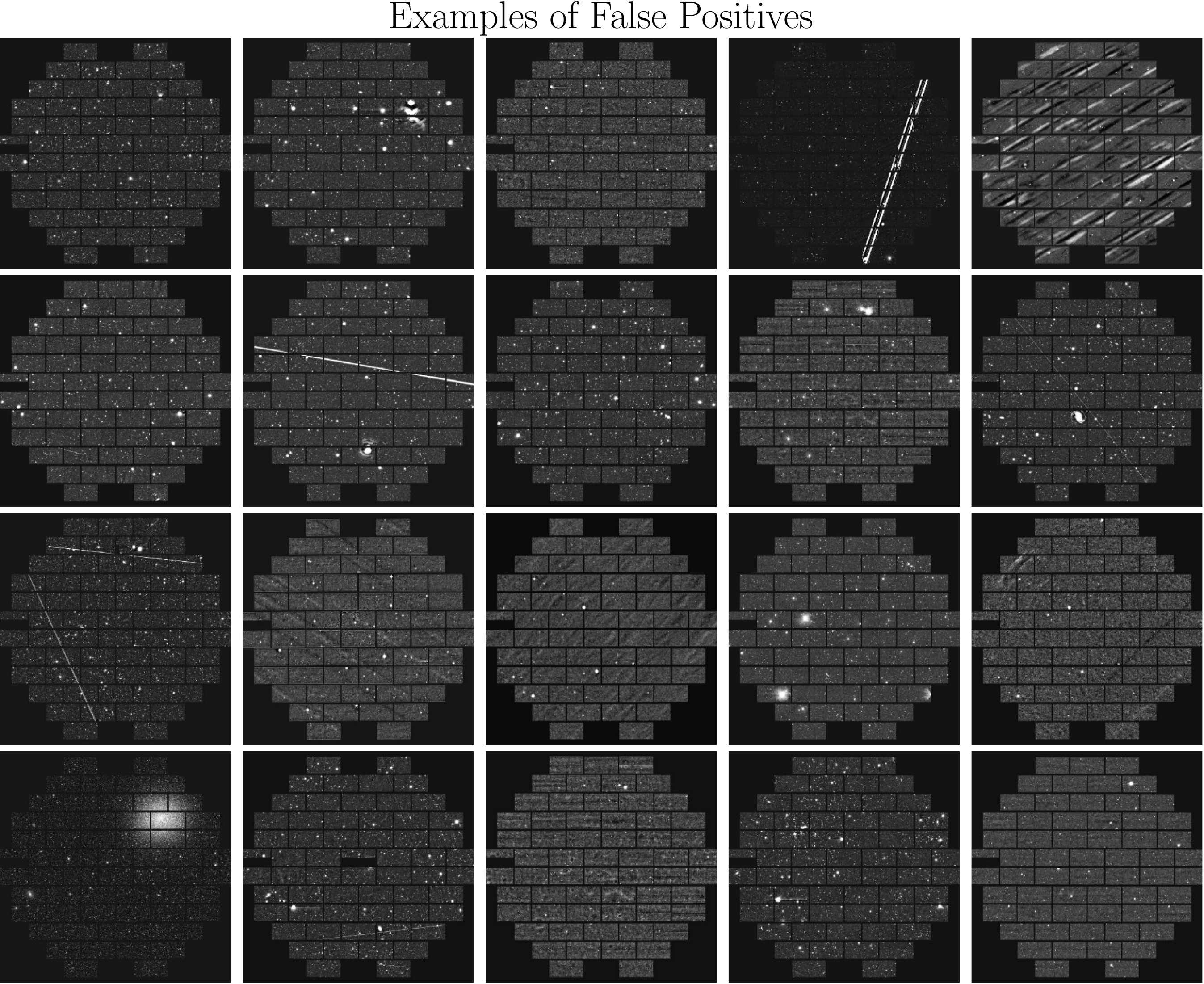}}
\subfigure[]{\includegraphics[width=0.8\textwidth]{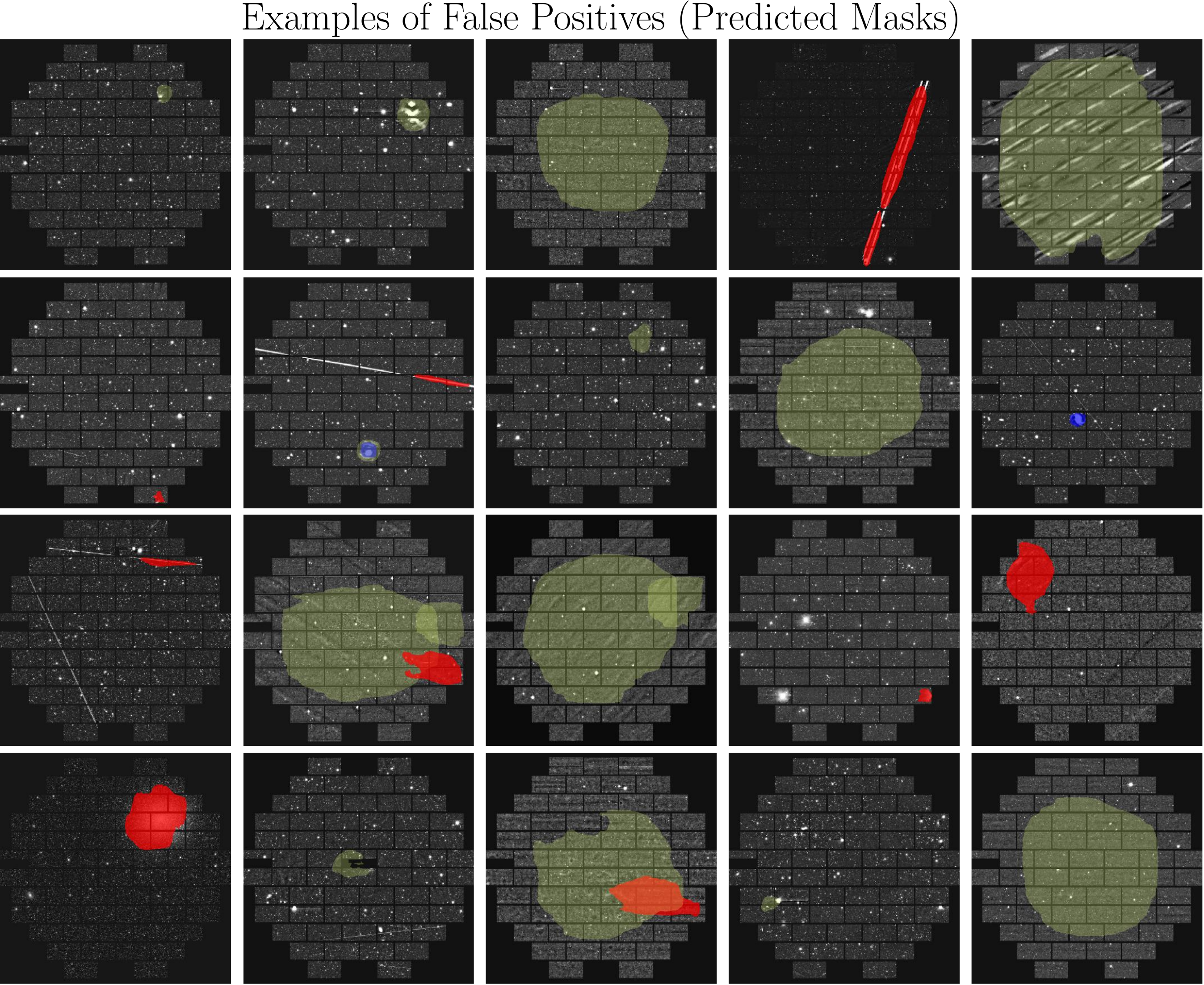}}
\caption{(a) Example images classified as ghost-containing (false positives) and the corresponding predicted masks (lower panel, (b)).}
\label{Fig: False_Positives}
\end{figure*}

Here we present examples of false positive and false negative classifications of ghosts and scattered-light artifacts from the Mask R-CNN method outlined in Sec.~\ref{sec: Classifier}.
Fig.~\ref{Fig: False_Positives} presents examples of false positives (panel (a)) and the corresponding mask predictions of the Mask R-CNN model (panel (b)) for the same images. The color scheme of the predicted masks follows that of the main text (see Fig.~\ref{fig: Prediction_Examples}).

As discussed in the main text, Sec. \ref{sec: Classifier}, most of those images are qualitatively different from other ghost-free images and contain either other types of artifacts --- for example, Earth-orbiting satellites ((2,2), (3,1)), airplane trails (1,4), structured cloud cover ((1,5), (3,2), (3,3)) or large galaxies ((2,2), (2,5)) and resolved stellar systems (4,1), \NEW{where the tuplets signify rows and columns, respectively.}

\begin{figure*}[!ht]
\centering
\includegraphics[width=0.8\textwidth]{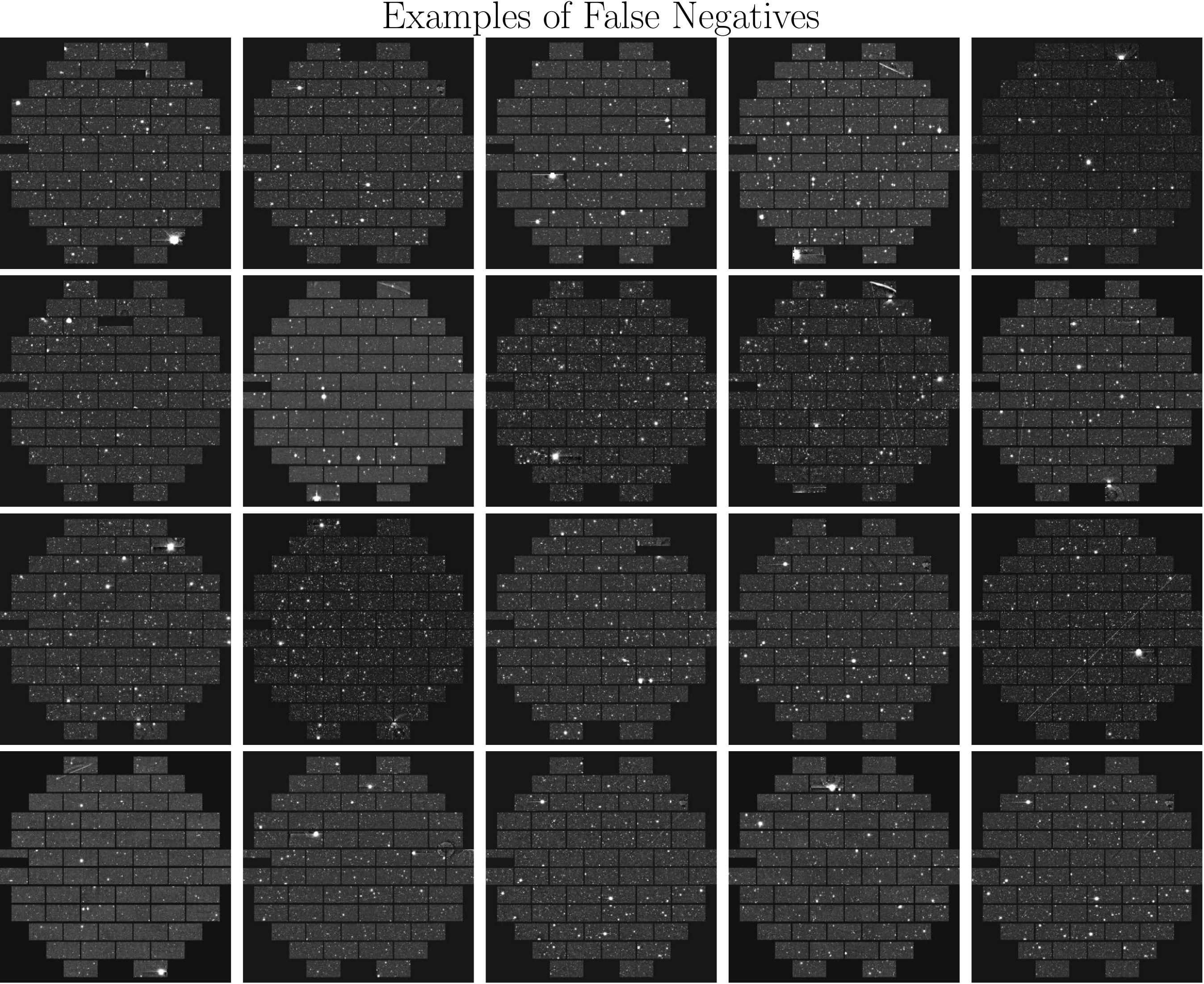}
\caption{Examples of false negatives, i.e. images that were classified as `clean' by the Mask R-CNN model (no objects detected). In practice, the artifacts present in these images are very small and faint, and often go undetected by human annotators.}
\label{Fig: False_Negatives}
\end{figure*}

Fig.~\ref{Fig: False_Negatives} presents some examples of false negatives. These images contain ghosts (as confirmed by visual inspection), but they are actually very small or faint and hard to distinguish at the resolution presented here. Thus, it is not a surprise that these have been classified as ``clean" by the mask R-CNN model, because they are different from the more prominent ghost-containing images that the network was trained on.

\bibliographystyle{model2-names}
\bibliography{main}







\end{document}